\documentclass[aps, pra, twocolumn, superscriptaddress, longbibliography, 10pt]{revtex4-2}
\usepackage[T1]{fontenc}
\usepackage[english]{babel}
\usepackage{amsmath}  
\usepackage{amsfonts} 
\usepackage{amssymb} 
\usepackage{amsthm}
\usepackage{mathtools}
\usepackage{physics}
\usepackage{bbold}
\usepackage{cancel}
\usepackage{booktabs}
\usepackage{enumitem}
\usepackage{comment}
\usepackage{graphicx}
\usepackage{footnote}
\usepackage{verbatim}
\usepackage{xcolor}
\usepackage{setspace}

\usepackage[colorlinks,bookmarks=false,citecolor=blue,linkcolor=red,urlcolor=blue]{hyperref}
\usepackage[capitalise]{cleveref}
\Crefname{figure}{Fig.}{Figs.}
\Crefname{section}{Sec.}{Secs.}
\usepackage{tipa}
\usepackage{wasysym}
\usepackage{bbold}
\usepackage[abs]{overpic}
\usepackage{stackengine, scalerel}

\def\cmp{\mathbin{\ThisStyle{\ensurestackMath{\abovebaseline[-\dimexpr1.1pt+0.55\LMpt]{%
  \stackunder[-\dimexpr1pt+2.5\LMpt]{\color{mygreen}\SavedStyle-}{%
  \color{red}\SavedStyle+}}}}}}

\graphicspath{{figures/}}
\usepackage[normalem]{ulem}
\makeatletter
\newcommand{\padA}{\affiliation{Dipartimento di Fisica e Astronomia ``G. Galilei'', Università di Padova, I-35131 Padova, Italy.}}
\newcommand{\padB}{\affiliation{Padua Quantum Technologies Research Center, Università degli Studi di Padova.}}
\newcommand{\padC}{\affiliation{Istituto Nazionale di Fisica Nucleare (INFN), Sezione di Padova, I-35131 Padova, Italy.}}
\newcommand{\padD}{\affiliation{Dipartimento di Fisica, Università di Bari, I-70126 Bari, Italy.}}
\usepackage{orcidlink}
\newcommand{\orcidgiovanni}{\orcidlink{0000-0002-9073-8978}}
\newcommand{\orcidgiuseppe}{\orcidlink{0000-0002-7280-445X}}
\newcommand{\orcidpietro}{\orcidlink{0000-0001-5279-7064}}
\newcommand{\orcidsimone}{\orcidlink{0000-0002-8882-2169}}
\definecolor{boxback}{HTML}{FFF8B5}
\definecolor{applegreen}{rgb}{0, 0.5, 0.0}
\definecolor{smoothred}{HTML}{C5232F}
\definecolor{mygreen}{rgb}{0,0.5,0}
\definecolor{myblue}{rgb}{0,0,0.75}
\definecolor{mymagenta}{cmyk}{0,1,0,0.12}

\newcommand{\red}[1]{{\textcolor{red}{#1}}}
\newcommand{\green}[1]{{\color{mygreen} #1}}
\newcommand{\avg}[1]{\left\langle#1\right\rangle}
\newcommand{\five}[5]{\ket{\scriptstyle{\begin{array}{ccc}
        & {#5} & \\
   {#2} & {#1} & {#4} \\
        & {#3} &
	\end{array} }}}

\begin{document}

\title{Simulating (2+1)D SU(2) Yang-Mills Lattice Gauge Theory\\ 
at finite density with tensor networks}
\author{Giovanni Cataldi\orcidgiovanni} \padA\padB\padC
\author{Giuseppe Magnifico\orcidgiuseppe} \padA\padB\padC\padD
\author{Pietro Silvi\orcidpietro} \padA\padB\padC
\author{Simone Montangero\orcidsimone} \padA\padB\padC

\date{\today}

\begin{abstract}
\noindent
We numerically simulate a non-Abelian lattice gauge theory in two spatial dimensions, with Tensor Networks (TN), up to intermediate sizes ($>$30 matter sites) well beyond exact diagonalization.
We focus on the SU(2) Yang-Mills model in Hamiltonian formulation, with dynamical matter and minimally truncated gauge field (hardcore gluon).
Thanks to the TN sign-problem-free approach, we characterize the phase diagram of the model at zero and finite baryon number as a function of the quark bare mass and color charge. 
At intermediate system sizes, we detect a liquid phase of quark-pair bound-state quasi-particles (baryons), whose mass is finite towards the continuum limit.
Interesting phenomena arise at the transition boundary where color-electric and color-magnetic terms are maximally frustrated: for low quark masses, we see traces of potential deconfinement, while for high masses, signatures of a possible topological order.
\end{abstract}

\maketitle

Non-Abelian gauge field theories, such as Quantum Chromodynamics (QCD), lay at the core
of the Standard Model of particle physics.
\begin{figure}[t]
\centering
\includegraphics[width=1\columnwidth]{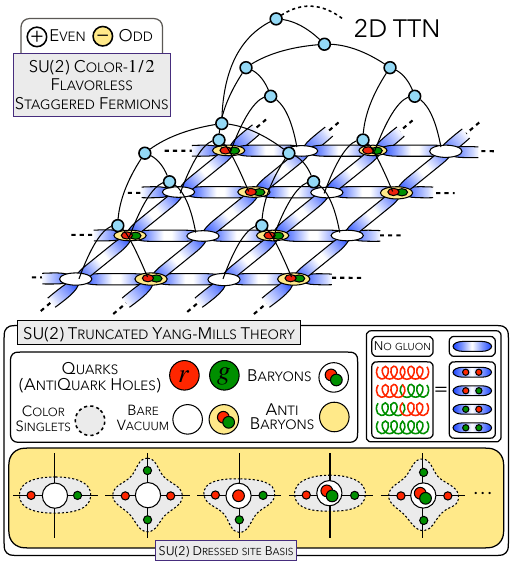}
\caption{TTN approach to (2+1)D hardcore SU(2) Yang-Mills LGT. Lattice sites host flavorless SU(2)-color-1/2 fermionic fields (red and green) in a staggered configuration (white and yellow). 
Lattice (blue) links describe gauge degrees of freedom from a 5-dimensional truncated Hilbert space. SU(2) Gauss Law is implemented at each lattice site.}
\label{fig_introduction_picture}
\end{figure}
They have been extensively successful in 
predicting the physical phenomena of quarks and gluons with large momentum transfers, where perturbative methods apply.
Conversely, at the energy scales of the hadronic world, where perturbative methods fail, robust numerical frameworks were developed,
such as lattice gauge theories (LGTs) \cite{Kogut1979, Rothe2012}. 
Monte Carlo (MC) simulations of LGTs characterized essential phenomena such as the hadronic spectrum, the mechanism for confinement \cite{Wilson1974, Kogut1983, Gupta1998}, the chiral symmetry breaking mechanism \cite{Kogut1982, Gottlieb1987, Alford1999, Mitter2015}, and the role of topology in QCD at finite temperatures \cite{Creutz1980, Creutz1982, Kogut1985, Biswal2017}.
Despite an impressive number of successful predictions, MC methods are hindered by the infamous sign problem, which hampers the simulation of a wide class of physical settings described by complex or negative actions (finite charge-density phases, fermions, real-time dynamics), whose numerical investigations remain -- to date -- an open challenge~\cite{Loh1990, Troyer2005}.     

In the last decade, following Feynman's seminal proposal and the recent fast development of quantum computers and simulators, quantum-inspired strategies attacked this challenge. On one hand side, atomic quantum simulators attempted to reproduce the quantum dynamics of lattice gauge theories \cite{Martinez2016, Schweizer2019, Yang2020, Zhou2022, Nguyen2022, Mildenberger2022}. 
On the other hand, Tensor Networks (TN) methods were identified as a powerful sign-problem-free numerical tool for complex lattice models \cite{Verstraete2008, Orus2019, Montangero2018, Silvi2019}. 
Exploiting TN algorithms, noteworthy results have been produced for Abelian gauge theories in (1+1)D \cite{Banuls2013, Rico2014, Kuhn2014, Banuls2015, Buyens2016, Buyens2017, Buyens2017a, Ercolessi2018, Magnifico2019, Magnifico2019a, Funcke2020, Magnifico2020, Rigobello2021, Banuls2022} and higher spatial dimensions \cite{Felser2020, Magnifico2021, Emonts2022}. 
As for non-Abelian gauge symmetries, TN-based simulations were so far limited to one spatial dimension \cite{Silvi2017, Silvi2019a, Kadam2023}.

In this work, we overcome such limitation: we present the TN simulations of a (2+1)D Hamiltonian analogous to a SU(2) Yang-Mills LGT, with flavorless fermionic matter.
The 2-colored quarks are discretized as staggered fermions on the sites of a square lattice, whereas the non-Abelian gauge fields live on the lattice bonds, undergoing a Kogut-Susskind dynamics \cite{Kogut1975, Kogut1979}.
Precisely, this study considers 
the smallest nontrivial electrically-truncated
$(0{\otimes} 0){\oplus}(\frac{1}{2}{\otimes}\frac{1}{2})$ representation of the SU(2) gauge field (see \cref{fig_introduction_picture}).
This \emph{`hardcore-gluon'} approximation keeps solely states of the gauge field generated from the bare vacuum with (at most) a single application of the parallel transporter operator.

We report numerical simulation results for the model above, using Tree Tensor Network (TTN) methods from small to intermediate system sizes, up to 32 matter sites. We stress that only up to 6 sites of the TTN methods can be carried out at maximum bond dimension, and thus equivalent to Exact Diagonalization (ED), due to the inherent complexity of the model.
We describe several regimes of the model at equilibrium, including finite baryon number density. 
The analysis of the ground state properties of the system, for lattice sizes up to $4 \times 8$ as performed here, due to the rich structure of the quantum degrees of freedom, would require $>$160 qubits to describe on a quantum computer. 
We characterize the model phase diagram by evaluating multiple observables, such as energy gaps, matter/antimatter and color-charge densities, and gauge field distributions.

TNs are based on controlled wave-function variational ansatzes exploiting the area-law entanglement bounds satisfied by locally interacting many-body quantum systems. 
Thus, they allow an efficient representation of the low-energy sectors contributing to the equilibrium properties and (low-entangled) time evolution \cite{Eisert2010}. 
TN methods do not suffer from the aforementioned sign problem \cite{Meurice2022}.
In this framework, ansatzes like Matrix Product States (MPS), Projected Entangled Pair States (PEPS), and Tree Tensor Networks (TTN) have found increasing applications for studying quantum many-body systems and LGTs \cite{Verstraete2008, Orus2019, Cataldi2021, Ferrari2022, Rico2014, Kuhn2014, Zohar2021, Meurice2022, Tagliacozzo2014}.
One main challenge for numerical and quantum simulations of gauge theories is the finite-dimensional encoding of the continuous gauge fields. 
A few recipes are known to achieve this reduction, from finite groups
\cite{Ercolessi2018, Magnifico2020, Haase2021} to fusion algebra deformation \cite{Zache2023}.
We adopt an energy-cutoff truncation strategy similar to a Quantum Link Model (QLM) \cite{Horn1981, Orland1990, Chandrasekharan1997, Brower1999, Tagliacozzo2014}, an approach already considered for practical quantum simulation of LGTs \cite{Byrnes2006, Mathis2020, Davoudi2020, Mazzola2021, Kan2021, Zohar2021a, Mariani2023, Pomarico2023, Bauer2023, Fontana2023}.
In this sense, the TN approach and the presented results could be used for benchmarking and validating current and future experimental implementations on quantum hardware \cite{Zohar2013, Banerjee2013, Wiese2013, Tagliacozzo2013, Mezzacapo2015, Banuls2020,  Banuls2020b, Atas2021, Klco2022, Meurice2022, Atas2023, Davoudi2023, Zache2023a} and to systematically identify the quantum advantage threshold~\cite{Zhou2020, Ayral2023}.

The manuscript is organized as follows: \cref{model} introduces the SU(2) Yang-Mills lattice Hamiltonian, and illustrates the dressed-site formalism \cite{Tagliacozzo2013, Silvi2014, Zohar2018, Zohar2019} we adopt, built on top of an energy-truncated Kogut-Susskind formulation \cite{Kogut1975}. 
In \cref{results}, we present ground-state numerical simulation results for the effective Hamiltonian.
In \cref{sec_conclusions}, our conclusions and outlook are presented. 
Finally, the appendices contain additional technical details of the theoretical mapping and the numerical simulation settings.  

\section{Model: Lattice SU(2) Yang-Mills}\label{model}
Using Tensor Network methods, we numerically simulate a Hamiltonian lattice-gauge model corresponding to the SU(2) Yang-Mills lattice gauge field theory at low energies. 
We place the fermionic matter on a finite $L_{x}\times L_{y}$ lattice $\Lambda$ and control the following parameters of the model: the quark bare mass $m_0$, the quark color charge $q_c$, the lattice spacing $a$, and the baryon number density $b$.
Sites and links are respectively identified by the couple ($\vb{j}, \vb*{\mu}$), where $\vb{j}=(j_{x},j_{y})$ is any 2D site, while $\vb*{\mu}$ is one of the two positive lattice unit vectors: $\vb*{\mu}_{x}=(1,0)$, $\vb*{\mu}_{y}=(0,1)$. 
Lattice sites are occupied by matter fields, which we represent with SU(2)-color staggered (Dirac) fermions $\hat{\psi}_{\vb{j}, \alpha}$ \cite{Susskind1977}, satisfying
\begin{align}
\qty{\hat{\psi}^{\dagger}_{\vb{j},\alpha}\hat{\psi}_{\vb{j'},\beta}}
    &=\delta_{\vb{j},\vb{j'}}\delta_{\alpha,\beta},&
  \text{where}&& 
  \alpha,\beta&\in\qty{\red{r},\green{g}}
\end{align}
are SU(2)-colors.
Then, the Hamiltonian reads:
\begin{equation}
  \begin{aligned}
    \hat{H}_0=&+
    \frac{c\hbar}{2a}\sum_{\alpha,\beta}\sum_{\vb{j}\in \Lambda} \Big[\text{-i} \hat{\psi}^{\dagger}_{\vb{j}, \alpha} \hat{U}_{\vb{j},\vb{j}+\vb*{\mu}_x}^{\alpha\beta} \hat{\psi}_{\vb{j}+\vb*{\mu}_x,\beta}\\
    & 
    - (-1)^{j_x + j_y}\hat{\psi}^{\dagger}_{\vb{j},\alpha} \hat{U}_{\vb{j},\vb{j}+\vb*{\mu}_y}^{\alpha\beta} \hat{\psi}_{\vb{j}+\vb*{\mu}_y,\beta} + \text{H.c.} \Big]\\
    &+ m_0 c^{2} \sum_{\vb{j}\in \Lambda} (-1)^{j_x + j_y}\sum_{\alpha} \hat{\psi}^{\dagger}_{\vb{j},\alpha} \hat{\psi}_{\vb{j},\alpha} + \hat{H}_{\text{pure}}
  \end{aligned}
  \label{eq_H_SU2_full}
\end{equation}
where $c$ is the speed of light, $\hbar$ is the Planck constant, and $a$ is the lattice spacing.
The first two terms describe fermion-hopping between nearest-neighboring sites along the $(\vb{j},\vb{j}+\vb*{\mu})$ lattice link. To enforce gauge symmetry, the hopping mechanism has to be mediated by the SU(2)-parallel transporter operator $U_{\vb{j},\vb{j}+\vb*{\mu}}^{\alpha\beta}$, acting on the gauge fields which live on the lattice links.
The latter term, or staggered mass, ensures that the fermion fields, at low energies and free theory, correctly describe a Dirac 4-spinor field with bare mass $m_0$ \cite{Susskind1977, Rothe2012, Zache2018}.

We employ the Kogut-Susskind formulation of gauge field dynamics \cite{Kogut1975} for the pure Hamiltonian $\hat{H}_{\text{pure}}$, due to its simplicity.
Namely, we have
\begin{equation}
  \begin{aligned}
    \hat{H}_{\text{pure}}=&+g^2\frac{c \hbar}{2a} \sum_{\vb{j}\in \Lambda}\qty( \hat{E}^2_{\vb{j},\vb{j}+\vb*{\mu}_x} + \hat{E}^2_{\vb{j},\vb{j}+\vb*{\mu}_y})\\
    &-  g^{-2} \frac{8c \hbar}{a}\sum_{\square\in \Lambda} \sum_{\substack{\alpha,\beta,\\\gamma,\delta}}\Re \qty(\begin{array}{ccc}
    \ulcorner& \hat{U}^{\dagger}_{\gamma \delta}& \urcorner\\
    \hat{U}^{\dagger}_{\delta\alpha}& & \hat{U}_{\beta\gamma}\\
    \llcorner& \hat{U}_{\alpha\beta}& \lrcorner\\
    \end{array}),
  \end{aligned}
  \label{eq_H_SU2_pure}
\end{equation}
where the coupling $g(q_c, a)$ is dimensionless, but scales nonetheless with the lattice spacing $a$ to ensure that the color charge $q_c$ of a quark stays finite in the continuum limit.
Namely, in $D$ spatial dimensions, it should scale as $g(q_c,a) \propto q_c a^{\frac{3-D}{2}}$ (see \cref{app_dimensional_analysis}), assuming that the SU(2) Yang-Mills theory in 2D is indeed super-renormalizable \cite{Hamer1985}.

As it is, $\hat{H}_{\text{pure}}$ in \cref{eq_H_SU2_pure} is already a frustrated quantum model even without fermion fields ($m_{0}{\to}\infty$).
The first term represents the SU(2)-electric energy density and corresponds to the Casimir operator on every link:
\begin{equation}
    \hat{E}^2_{\vb{j},\vb{j}+\vb*{\mu}} = \abs{\hat{\vb{L}}_{\vb{j},\vb{j}+\vb*{\mu}}}^2 = \abs{\hat{\vb{R}}_{\vb{j},\vb{j}+\vb*{\mu}}}^2, 
    \label{Casimir}
\end{equation}
where $\hat{\vb{L}}_{\vb{j},\vb{j}+\vb*{\mu}}$ (resp. $\hat{\vb{R}}_{\vb{j},\vb{j}+\vb*{\mu}}$) are the group generators of the left (right) gauge transformations on the link, hermitian and satisfying, $\forall k\in\qty{x,y,z}$:
\begin{equation}
\begin{aligned}
    \qty[\hat{L}^{k},\hat{R}^{k'}]&=0\\
    \qty[\hat{L}^{k}_{\vb{j},\vb{j}+\vb*{\mu}},\hat{L}^{k'}_{\vb{j'},\vb{j'}+\vb*{\mu}'}]&=
    i \delta_{\vb{j} \vb{j'}} \delta_{\vb*{\mu} \vb*{\mu}'}
    \epsilon_{k k'}^{k''} {L}^{k''}_{\vb{j},\vb{j}+\vb*{\mu}},
\end{aligned}
    \label{Genalgebra}
\end{equation}
(same with $\hat{\vb{R}}$) with $\epsilon$ the Levi-Civita symbol.
The second contribution to \cref{eq_H_SU2_pure} approximates the SU(2)-magnetic energy density through the smallest Wilson loops, i.e. square gauge-invariant plaquettes made out of parallel transporters $\hat{U}$. 

According to Wilson’s formulation of LGTs, faithful representations of the local gauge field algebra satisfy
\begin{equation}
\begin{aligned}
    [\hat{L}^{k}_{\vb{j},\vb{j}+\vb*{\mu}},\hat{U}_{\vb{j}',\vb{j}'+\vb*{\mu}'}^{\alpha\beta}]&= - \delta_{\vb{j} \vb{j'}} \delta_{\vb*{\mu} \vb*{\mu}'} \sum_{\gamma} \frac{{\sigma}^{k}_{\alpha\gamma}}{2} \hat{U}_{\vb{j},\vb{j}+\vb*{\mu}}^{\gamma\beta},
    \\
    [\hat{R}^{k}_{\vb{j},\vb{j}+\vb*{\mu}},\hat{U}_{\vb{j}',\vb{j}'+\vb*{\mu}'}^{\alpha\beta}]&= + \delta_{\vb{j} \vb{j'}} \delta_{\vb*{\mu} \vb*{\mu}'} \sum_{\gamma}  \hat{U}_{\vb{j},\vb{j}+\vb*{\mu}}^{\alpha \gamma}    \frac{{\sigma}^{k}_{ \gamma \beta}}{2}
\end{aligned}
    \label{Gaugealgebra}
\end{equation}
for ${\sigma}^{k}$ Pauli matrices and $\hat{U}$ operators rescaled such that closed Wilson loops preserve the state norm.

To perform numerical simulations of the Hamiltonians in \cref{eq_H_SU2_full}-\eqref{eq_H_SU2_pure}, we need to achieve a finite yet controllable truncation of the local gauge Hilbert space. 
As detailed in \cref{app_model}, we develop an energy-cutoff truncation strategy that is similar to the Quantum Link Model (QLM) \cite{Chandrasekharan1997}, an approach that has been already adopted for quantum simulation of LGTs \cite{Byrnes2006, Mathis2020, Davoudi2020, Mazzola2021, Kan2021, Zohar2021a, Mariani2023, Pomarico2023, Bauer2023}. 
Our formalism is self-consistent, scalable to arbitrarily large truncations, and applicable to lattices of any spatial dimension.
Nonetheless, all the results of this work refer to the smallest non-trivial energetic truncation, which we label as \emph{hardcore-gluon approximation}.

\subsection{Hardcore-gluon approximation}
In analogy to cold quantum gases, the label \emph{hardcore-gluon} aims to stress that the only accessible local configurations are those states reachable from the bare vacuum with at most one application of $\hat{U}$.
Namely, we consider $(0{\otimes} 0){\oplus}(\frac{1}{2}{\otimes}\frac{1}{2})$ as the gauge field space (dimension 5), where $(s)$ is the irreducible spin-$s$ representation of SU(2) \cite{Horn1981, Orland1990, Brower1999}. 
This is the smallest representation ensuring a nontrivial contribution of all the terms in the Hamiltonian \cref{eq_H_SU2_full}-\eqref{eq_H_SU2_pure}. 
The truncation keeps the electric field operator $\hat{E}$ hermitian and protects the algebra rules of \cref{Gaugealgebra}, but $\hat{U}$ is no longer unitary (it loses norm on the largest spin shell). 
Moreover, it introduces a local energy cutoff in units of 
$g^2 a^{-1}\propto q_c^2$. 
This is the scaling, as a function of $a$, of the electric energy coupling and the bare mass energy coupling.

To accurately represent the full theory, for weak-$g$, larger gauge representations are required: this increases the computational challenges 
but it is still potentially accessible via TNs. 
In \cref{app_model}, we discuss in detail how to extend the effective model to arbitrary truncation for spin-shells, in a practical way that can be readily implemented with TNs or in an analog/digital quantum simulation.
As a final step of the mapping, we define an effective  Hamiltonian 
(also discussed in \cref{app_model}) 
which acts on logical sites built merging gauge and matter degrees of freedom in a compact dressed-site formalism \cite{Tagliacozzo2013, Silvi2014, Zohar2018}. 
Correspondingly, as done in Loop String Hadrons methods \cite{Raychowdhury2020}, the original non-Abelian gauge invariance is exactly rewritten into an Abelian, nearest-neighbor, diagonal selection rule, and the explicit dependence on the fermionic matter is eliminated \cite{Felser2020, Zohar2018a, Zohar2019}.

We also stress that large-$g$ regime can be addressed by exploiting perturbation theory in $1/g^2$ (carried out in \cref{app_large_g_perturbation_theory}).
In this scenario, the full theory can be mapped to a good approximation into a spin-like Hamiltonian similar to a 2D anisotropic Heisenberg model \cite{Wang1991, Wiese1994}.

\section{Results}
\label{results}
This section collects the numerical results from the ground states of SU(2) Hamiltonian in \cref{eq_H_SU2_full}-\eqref{eq_H_SU2_pure}, obtained via Tree Tensor Network simulations (TTN), for small (maximum bond dimension, i.e. ED) and intermediate system sizes.
Hereafter, we rescale the Hamiltonian in dimensionless energy scale units
$\hat{H}_0 {\to} \hat{H} {=} \frac{a}{c \hbar} \hat{H}_0$ so that the hopping term has constant coupling $\frac{1}{2}$. 
Correspondingly, the other Hamiltonian terms acquire the re-scaled dimensionless couplings $m{=}m_0 a c/\hbar {=} (a/a_m)$ (staggered mass), $\frac{1}{2} g^2 {=}\frac{1}{2} (a/a_g)$ (electric) and $8 g^{-2}{=}8 (a_g/a)$ (magnetic), where we considered $g$ to scale as $g{\propto} a^{1/2}$ in two spatial dimensions (see \cref{app_dimensional_analysis}). 

If we exclude quantum corrections to the scaling (anomalous dimension \cite{Giedt2016}), then the continuum limit is located at $g^2{=}\alpha_c{m}{\to}0$ (more quantitatively at $a{\ll}a_g, a_m$). 
The fixed dimensionless \emph{'quark ratio'} $\alpha_c {=} g^2/2m {=}\frac{1}{2} (a_m/a_g)$ does not scale with the lattice spacing and is solely determined by the color charge and the bare mass of the quark  (see \cref{app_dimensional_analysis}).

Together with the ground-state energy density $\varepsilon=\langle\hat{H}\rangle/\abs{\Lambda}$, we evaluate the expectation values $\avg{\cdot}$ of several local observables onto the computed ground states. 
Regarding gauge fields, we track the color-electric and color-magnetic energy densities
\begin{align}
  \avg{E^{2}}&=\frac{1}{\abs{\Lambda}}\sum_{\vb{j}\in \Lambda}\sum_{\vb*{\mu}}\avg{\hat{E}^{2}_{\vb{j},\vb*{\mu}}}\label{gauge_observables1}\\
  \avg{B^{2}}&=-\frac{1}{\abs{ \square}}\sum_{\square\in\Lambda}
  \Re\avg{\begin{array}{ccc}
    \ulcorner& \hat{U}^{\dagger}& \urcorner\\
    \hat{U}^{\dagger}& & \hat{U}\\
    \llcorner& \hat{U}& \lrcorner\\
    \end{array}}+c',
\label{gauge_observables2}
\end{align}
where $\abs{\Lambda}$ and $\abs{\square}$ correspond to the total number of sites and lattice plaquettes. The constant factor $c'=\frac{1}{2}$ in \cref{gauge_observables2} sets the minimum of the magnetic energy density to 0.
When considering the matter, it is useful to separately measure the staggered fermion density for even (+) and odd (-) sites
\begin{align}
    N_{\pm}&=
    \frac{1}{\abs{\Lambda_{\pm}}}\sum_{\vb{j}\in 
    \Lambda_{\pm}} \sum_{\alpha=\red{r},\green{g}}    \langle \hat{\psi}^{\dagger}_{\vb{j},a} \hat{\psi}_{\vb{j},a}
    \rangle
    \label{avg_particle_density}
\end{align}
where $\Lambda_{+}$ (resp. $\Lambda_{-}$) is the even (odd) sub-lattice. 
Tracking these two quantities gives us immediate access to the density of quarks $(N_+)$ and the density of anti-quarks $(2-N_{-})$ separately, according to the staggered fermion formalism.
Similarly, we can define
the total \emph{particle density} (quarks plus anti-quarks)
\begin{align}
\varrho &= N_{+} + ( 2 - N_{-} )&
0&\leq\varrho\leq4
\label{density}
\end{align}
as well as the \emph{baryon number density}, (quarks minus anti-quarks divided by two)
\begin{align}
b &= \frac{1}{2}(N_{+} - (2 - N_{-}))
&
0&\leq b\leq1
\end{align}
which is a good quantum number, as it is a conserved quantity tied to the global staggered fermion number conservation.
We stress that, unlike quantum chromodynamics, SU(2) Yang-Mills baryons $-$ colorless bound states of matter particles $-$ are made by two, not three, quarks. 
Similarly, anti-baryons are made by two anti-quarks. Correspondingly, mesons are made by one quark and one anti-quark as normal.

Both mesons and standalone quarks can be detected by looking at the average \emph{matter color density} $|\vb{S}|^{2}$, that is, the quadratic Casimir operator of the matter field gauge group transformations:
\begin{equation}
  |\vb{S}|^{2}{=} \frac{1}{\abs{\Lambda}}\sum_{\vb{j}}\avg{\hat{\vb{S}}^{2}_{\vb{j}}}{=}\frac{1}{2\abs{\Lambda}}\sum_{\vb{j},\alpha\beta}\avg{\qty(\hat{\psi}^{\dagger}_{\vb{j},\alpha}\hat{\psi}_{\vb{j},\beta}\vb*{\sigma}_{\alpha\beta})^{2}},
  \label{eq_matter_Casimir}
\end{equation}
where $\alpha,\beta \in \qty{\red{r},\green{g}}$.
Our quantitative analysis also includes the von Neumann entanglement entropy \cite{Eisert2010}
\begin{equation}
  \mathcal{S}_{A}=-\Tr \rho_{A}\log_{2}\rho_{A},
  \label{entropy}
\end{equation}
where $\rho_{A}$ is the reduced density matrix of the partition $A$, which we choose exactly to be the bottom (or top) half of the system.

\subsection{Magneto-electric transition in the pure theory}\label{sec_pure_theory}
\begin{figure}
	\centering
	\includegraphics[width=1\columnwidth]{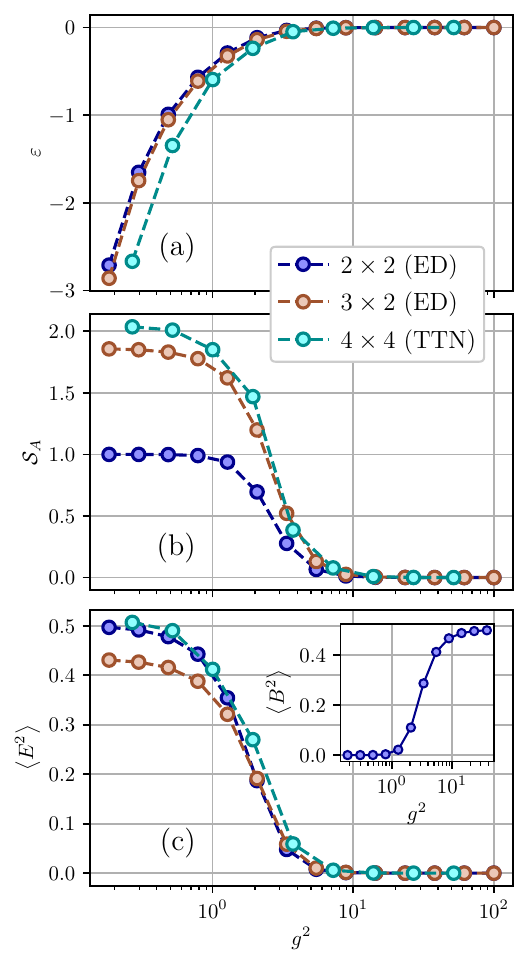}
	\caption{
            Numerical simulations of the pure Hamiltonian in \cref{eq_H_SU2_pure} with OBC for different lattice sizes. 
            The plots display respectively (a) the ground-state energy density $\varepsilon$, (b) the entanglement entropy $\mathcal{S}_{A}$ of half the system, (c) the average electric energy contribution $\avg{E^{2}}$, with the magnetic energy density $\avg{B^{2}}$ shown in the inset.}
	\label{pure_theory_simulations}
\end{figure}

We first focus on the pure theory ($m = \infty$) under Open Boundary Conditions (OBC). 
According to the results shown in \cref{pure_theory_simulations}, the pure Hamiltonian displays two phases driven by $g$. 
In the small-$g$ (magnetic) phase, the plaquette interactions provide the larger contribution to the energy in \cref{eq_H_SU2_pure}. 
As such, magnetic fields are depleted, and electric fields display large quantum fluctuations (see \cref{app_ED_results}) and compensate for any electric activity. 
Conversely, in the large-$g$ (electric) phase, electric fields are energetically expensive and thus depleted in the ground state, while magnetic fields show large fluctuations.

Unlike the electric phase, which displays marginal entanglement as the ground state is almost a product state, the magnetic phase reveals an entanglement that scales with the length of the bi-partition: 
this behavior, signaling a sharp area-law of entanglement, suggests that the magnetic phase is likely approximated by a resonant-valence bond state of plaquettes, akin to the local structure of the ground state of the Toric Code \cite{Kitaev2006a}.

The entanglement entropy approximates a monotonic function along $g$, without any peak in the transition between the two phases.
This observation suggests that, for large bare masses $m$, this quantum phase transition is either \emph{first order} or a \emph{crossover}.
Conversely, as shown in \cref{fig_ED_SU2_grid_discrete}, the small-$m$ scenario of the full theory peaks close to the transition, and the peak is wider and larger for smaller masses. 
We stress that the magneto-electric transition is compatible with the \emph{roughening transition} \cite{Drouffe1981, Munster1981, Kogut1983} observed via MC simulations \cite{Berg1981, Ambjorn1984a} and Cluster Expansion Methods (CEM) \cite{Hamer1985, Arisue1984}.

\subsection{Baryonic spectrum}
\label{res_energy_gaps}
For finite $m$, fermionic matter is included in the full Hamiltonian of \cref{eq_H_SU2_full}. 
The baryon number density $b$ is a quantum number associated with global symmetry, and can thus be directly encoded in the TTN ansatz. 
In this way, we directly target the ground state within a selected baryon number density sector \cite{Silvi2014, Silvi2019}.

The model is symmetric under CP, that is, mirror spatial reflection ($j_x \to L_x - j_x$) times particle-hole exchange
(${\hat{\psi}}_{\alpha} \to
i \sigma^{y}_{\alpha\beta} {\hat{\psi}}_{\beta}^{\dagger}$) of staggered fermions. 
Then, at negative baryon densities $b{<}0$, the ground state is the CP-reflected of the ground state at positive baryon density $|b|$.

We numerically verified that the global ground state is found at null baryon density $b=0$ for any $g$ and $m$. As we can directly tune the baryon number of each TTN simulation, we have immediate access to the inter-sector energy gap by calculating the difference
\begin{equation}
\begin{split}
    \Delta_{\abs{b}} &= \qty(\varepsilon_{b} - \varepsilon_{0})\abs{\Lambda} =
    \qty(\varepsilon_{-b} - \varepsilon_{0})\abs{\Lambda}
    \geq 0\\
    &=m\abs{b}\abs{\Lambda}+ \Delta_{\abs{b}}^{*},
\end{split}
\label{intersector_gap}
\end{equation}
where we also defined the \emph{binding energy} $\Delta_{\abs{b}}^{*}$ by subtracting the bare mass of the corresponding excess quarks or anti-quarks ($\abs{b}\abs{\Lambda}$).

A simple yet illustrative analysis is to study the energy density gap between the one-baryon sector ($b{=}2/\abs{\Lambda}$) and the vacuum sector ($b{=}0$) and then approach the continuum limit $a\to0$ at fixed ratio $\alpha_c {=} g^2/m \propto q_c^2 / m_0$.

\begin{figure}
	\centering
	\includegraphics[width=1\columnwidth]{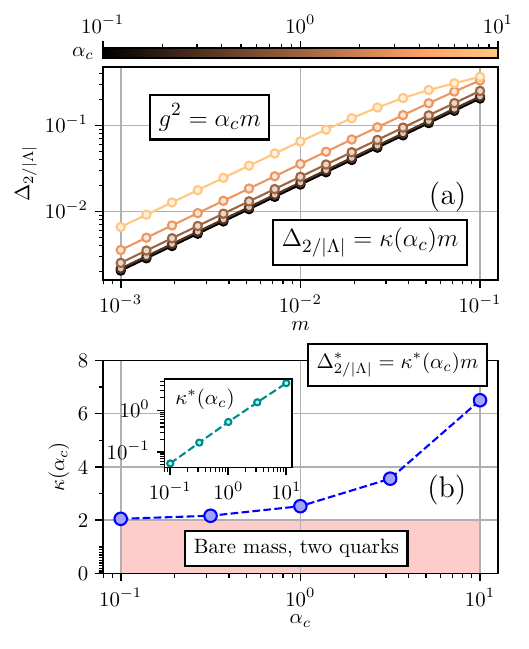}
    \caption{
        (a) Scaling of the inter-sector gap $\Delta_{2/\abs{\Lambda}}$ in \cref{intersector_gap} as a function of $m$, for different choices of the $g$-coupling $g^{2}=\alpha_{c}m$. 
        By fitting the power-law scaling of $\Delta_{2/\abs{\Lambda}}$ in the small-$m$ limit, we obtain the linear dependence on $m$ shown in \cref{baryon_mass}, whose slope $\kappa$ depends on $\alpha_{c}$ as shown in (b). The inset displays the corresponding $k^{*}$ of  the binding energy $\Delta_{2/\abs{\Lambda}}^{*}$ in \cref{binding_energy}.
        Results have been obtained from simulations of a $2\times 2$ lattice in PBC, where $\Delta_{2/\abs{\Lambda}}=\Delta_{\abs{b}=0.5}$.
    }
    \label{fig_energy_gaps} 
\end{figure}

As shown in \cref{fig_energy_gaps}(a), the gap $\Delta_{2/\abs{\Lambda}}$ displays a clear linear scaling with $m = \frac{m_0 c}{\hbar} a$.
Namely, we obtain:
\begin{align}
    \Delta_{2/\abs{\Lambda}} &=\kappa(\alpha_{c})m
    = \frac{m_0 c}{\hbar} \kappa(\alpha_{c}) a,
    \label{baryon_mass}
\end{align}
implying that the actual baryon mass is $m_{b}= \kappa(\alpha_{c}) m_{0}$.
As for all hadrons, its mass is always greater than the bare mass of its quark components, thus $\kappa \geq 2$. We show this observation in \cref{fig_energy_gaps}(b), where we display $\kappa$ as a function of $\alpha_{c}$.
More interestingly, in the case of the binding energy $\Delta_{2/\abs{\Lambda}}^*$ (inset of \cref{fig_energy_gaps}(b)), we observe a power-law scaling of $\kappa^*$ in $\alpha_{c}$:
\begin{align}
    \kappa^*&{=}\frac{\Delta_{2/\abs{\Lambda}}^*}{m}{=} \kappa{-}2 &
    \text{with}&&
    \kappa^* (\alpha_c)&{\sim} 0.13 \cdot \alpha_c^{0.96}
    \label{binding_energy}
\end{align}
compatible with linear scaling.
Such relations confirm that baryons are actual quasi-particles of the continuum theory and provide a connection to the bare quark properties ($\alpha_c$, $m_0$). 
We carried out this analysis for a finite-size sample, but the baryon-to-quark mass ratio $\kappa$ is expected to stay finite even at the thermodynamical limit.

\subsection{Baryon-liquid phase}
\label{sec_full_theory}
\begin{figure*}
    \centering
    \includegraphics[width=2\columnwidth]{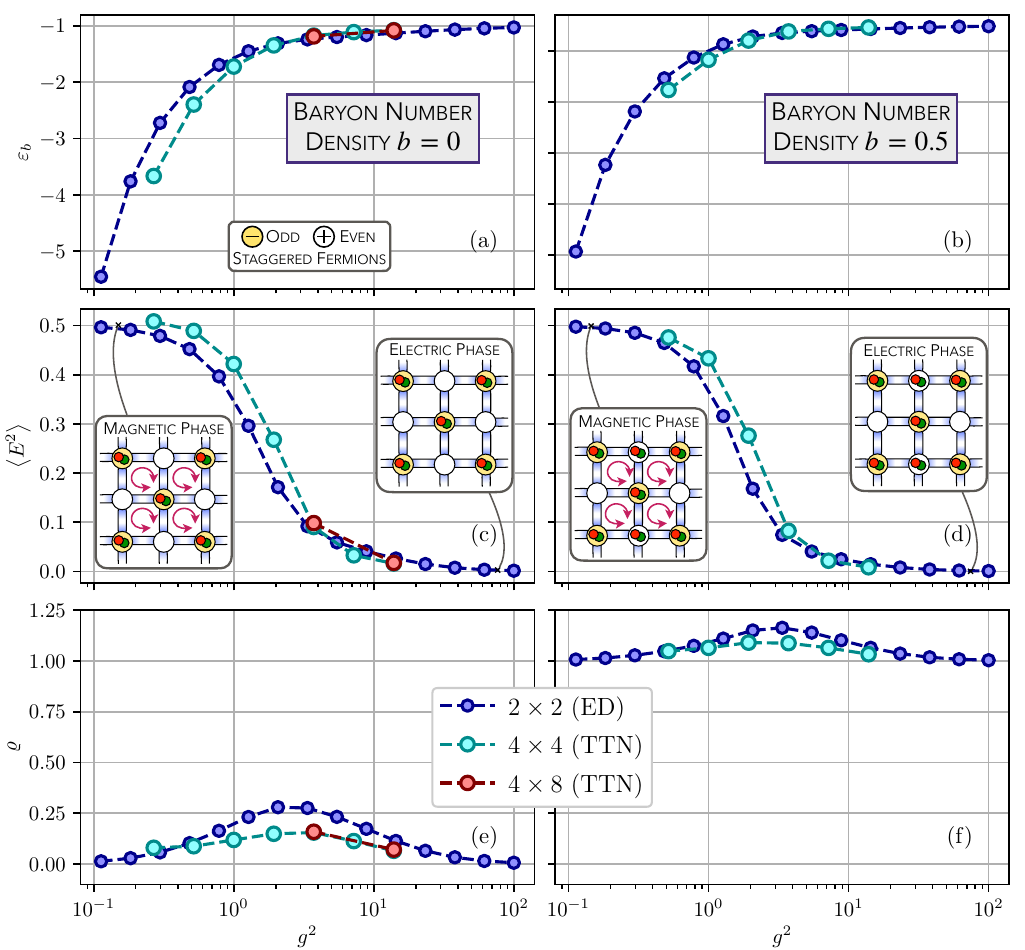}
    \caption{
    Numerical results of the full SU(2) Hamiltonian in \cref{eq_H_SU2_full} with OBC and baryon number density $b=0$ (left column) and $b=0.5$ (right column). The plots display respectively: 
    (a)-(b) the ground-state energy density $\varepsilon_{b}$, (c)-(d) the average electric energy contribution $\avg{E^{2}}$ in \cref{gauge_observables1}, enlightening the transition between the magnetic (purple fluxes) and the electric (no fluxes) phases discussed in \cref{sec_pure_theory}, and (e)-(f) the average particle density $\varrho$ in \cref{density}, which appears peaked in the $g$-transition. 
    The pictorial lattice configurations in the finite baryon density $b=0.5$ represent states with $b$ extra gapped hardcore local bosons with low dynamics compatible with the two electric/magnetic phases.
    }
    \label{full_ED_TTN}
\end{figure*}

Beyond energy gaps, other phase properties can be inferred when probing the observables in 
\cref{gauge_observables1}-\eqref{entropy}. 
The magneto-electric transition, driven by $g^2$, remains unaltered for finite $m$ and even at finite baryon densities $b$, as shown in \cref{full_ED_TTN}.

By contrast, the particle density $\varrho$ reveals an exciting behavior as the re-scaled quark mass $m$ is lowered. 
As long as $m$ is the largest energy scale of the model ($m \gg 1,g^2,g^-2$) the emergent behavior is relatively trivial, as a system of gapped hardcore bosons. 
More precisely, if $b \geq 0$ (resp.  $b \leq 0$) the antimatter (matter) sites are fully emptied, while the matter (antimatter) sites host exactly $b$ quark-pair hardcore bosons, mass gapped and with almost flat-band dynamics. 
The particle density $\varrho$ confirms this interpretation, as it stays at its minimum possible value of $\varrho \simeq \varrho_{\text{min}}(b) = 2|b|$ and having no fluctuations $\delta\rho \simeq 0$
(see for instance \cref{app_ED_results}).

\begin{figure}
	\centering
	\includegraphics[width=1\columnwidth]{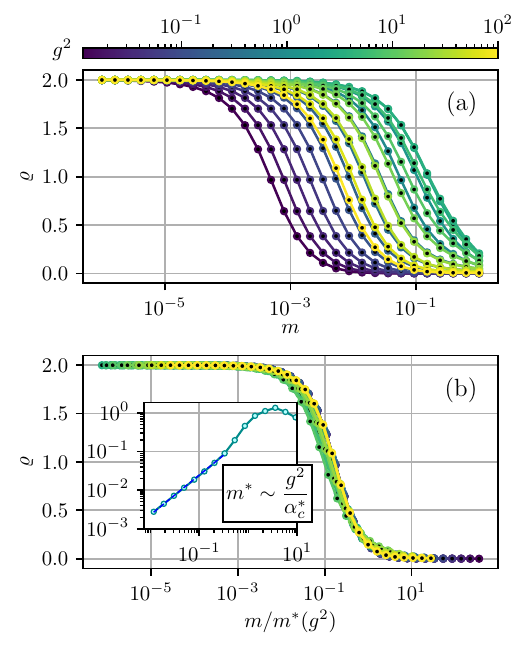}
    \caption{(a) Scaling of the particle density defined in \cref{density} as a function of $m$ for different values of the gauge coupling $g$. 
    (b) All the $\rho(m)$ curves of the particle density collapse on a single one simply by re-scaling the mass $m$ by a factor $m^{*}$ displaying a power-law scaling in $g^{2}$ (see the inset). 
    By fitting this scaling we extract \cref{eq_powerlaw_scaling}, whose error bars have been computed exploiting error propagation onto the covariance matrix of the fit.
    Results obtained from simulations on a $2\times 2$ lattice in OBC at baryon density $b=0$.}
    \label{fig_local_matter_density}
\end{figure}

The behavior drastically changes at low masses $m$, in relative proximity of the transition line $g^2 \sim 2(1)$, as shown in \cref{fig_local_matter_density}. 
In fact, for $m$ lower than a critical value $m^{*}(g)$, we see a sharp growth of the particle density $\rho$ and its on-site fluctuations $\delta \rho$, which become similar in magnitude (see \cref{app_ED_results}). 
Even though we do not have access to long-range correlation functions at these limited system sizes, this observation is a strong hint of superfluidity of the phase, where we expect the quasi-particle excitations to be gapless (in the rescaled units).

To deeper investigate the nature of these quasi-particles we track the matter-color density $|\vb{S}|^{2}$ (see \cref{fig_phase_diagram} and \cref{app_ED_results}).
There is a very narrow region around the magneto-electric transition where colored matter emerges (maybe a possible deconfined critical boundary). Elsewhere, especially towards the continuum limit, the color density stays $|\vb{S}|^{2} = 0$. 
We must conclude that the gapless quasi-particles must be made by on-site pairs of quarks or anti-quarks. As such, we can regard the low-mass phase, $m < m^{*}(g)$, as a {\it gapless baryon liquid}.

Using a finite-size scaling technique 
(shown in \cref{fig_local_matter_density}(b))
we are able to characterize $m^{*}$ as a power-law function of $g^2$, where a numerical regression yields
\begin{equation}
  m^{*}(g^{2}) \simeq 0.267(4) \cdot \qty(g^{2})^{1.03(2)},
  \label{eq_powerlaw_scaling}
\end{equation}
which is less than $2\sigma$ deviation from a linear scaling.
Suppose we now assume that the linear scaling holds, then there must be a critical quark ratio $\alpha^{*}_c = 3.75(6)$ that determines the behavior when approaching the continuum limit (recall that $\alpha^{*}_c$ depends only on quark color-charge and bare mass, see \cref{app_dimensional_analysis}). 
Namely, for strong color charges $\alpha_c > \alpha^{*}_c$ the the baryon fluid at $a \to 0$ is gapless, while for weak charges $\alpha_c < \alpha^{*}_c$ the baryon fluid is gapped.
We recall that we are working with energy scales rescaled by $a$, thus only quasiparticles that we identify as gapless at the continuum limit will survive as finite energy excitations in natural units.

\subsection{Non-local/Topological properties}\label{res_topology}
A relevant analysis that can be carried out in Yang-Mills theories is the characterization of topological properties at the critical point, and the investigation of whether some form of topological order emerges within or without deconfined phases \cite{Svetitsky1982, Tagliacozzo2011}. 
While the simplified model we considered does not support the existence of a deconfined phase in proximity to the continuum limit, it is still possible to characterize some topological properties by evaluating non-local order parameters.
As detailed in \cref{app_topology}, the pure theory in \cref{eq_H_SU2_pure} 
protects a topological symmetry, which exists only under periodic boundary conditions. Such symmetry is identified by the topological invariants (string operators) $\mathbb{P}_{x,y}$ defined in \cref{Py}-\eqref{Px} and forming a $\mathbb{Z}_2{\times}\mathbb{Z}_2$ group.

\begin{figure}
    \centering
    \includegraphics
    [width=1\columnwidth]
    {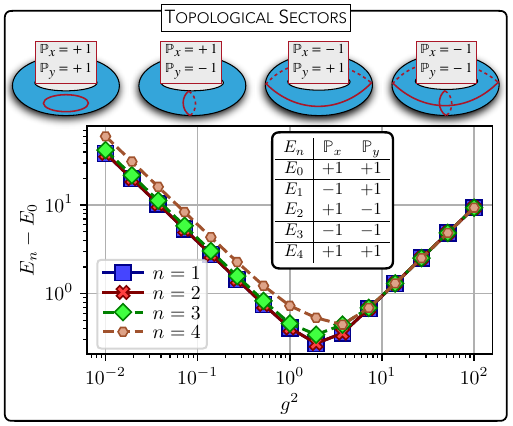}
    \caption{
    Energy gaps between the first excited levels and the ground state of \cref{eq_H_SU2_pure} in PBC, for a $2\times 2$ lattice. 
    Every state belongs to one of the topological sectors sketched on top: closed red curves on the blue torus correspond to SU(2) electric-loop excitations.}
    \label{fig_topological_sectors}
\end{figure}

By selecting each quantum number(s) for this symmetry group, we can evaluate inter-sector and intra-sector energy gaps, and verify the presence of quasi-degeneracies, signatures of a potential spontaneous breaking of the topological symmetry group, and thus of topological order.
As shown in \cref{fig_topological_sectors}, when approaching the transition point from the large-$g$ phase, inter-sector and intra-sector gaps reach a minimum, signaling a possible degeneracy lifted by finite-size effects. 
However, both gaps re-open while moving towards the small-$g$ phase. 
This observation suggests topological order not to survive for $g^2 \ll 2$.

The addition of dynamical matter removes the topological invariants $\mathbb{P}_{x,y}$ from being symmetries of the model, due to the hopping term inverting the string parity (see \cref{fig_full_topology}).
In the large-$m$ limit, where the particle density vanishes, the full theory approaches the pure one, and the topological invariants become good quantum numbers again.

\subsection{(2+1)D SU(2) Yang-Mills LGT Phase Diagram}
\label{sec_phase_diagram}
\begin{figure}
	\centering
	\includegraphics[width=1\columnwidth]{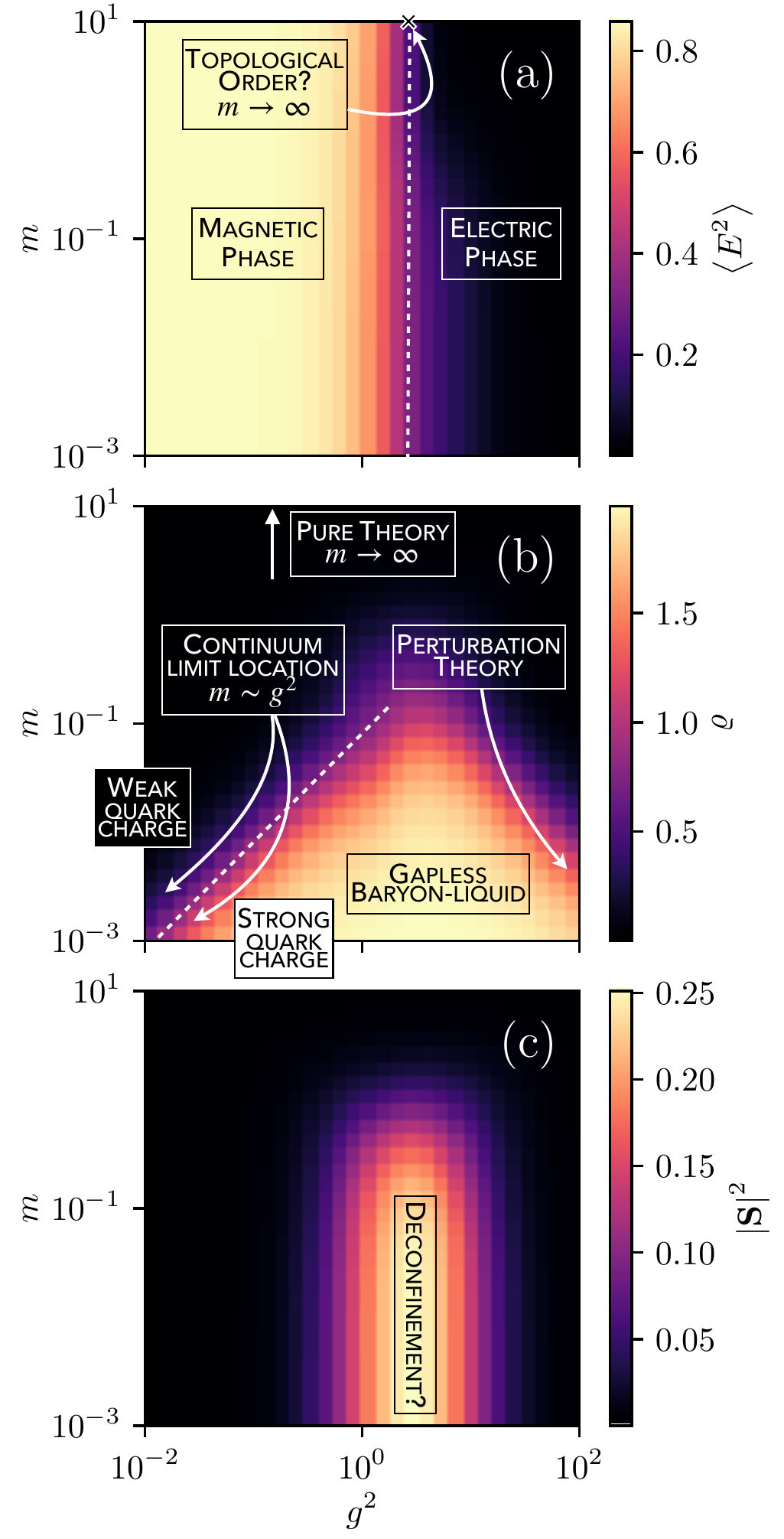}
 \caption{Phase diagram $(g^{2},m)$ of the full SU(2) Hamiltonian in \cref{eq_H_SU2_full}-\eqref{eq_H_SU2_pure} in the sector with zero baryon number density from (a) the average electric energy density in \cref{gauge_observables1}, (b) the average particle density in \cref{density}, and (c) the matter color density defined in \cref{eq_matter_Casimir}. 
 Phases are marked according to the discussion in \cref{results}, and \cref{app_ED_results}.}
\label{fig_phase_diagram}
\end{figure}

By collecting all the previous observations, we can outline in \cref{fig_phase_diagram} the full phase diagram of the 2D SU(2) Yang-Mills Hamiltonian in \cref{eq_H_SU2_full}-\eqref{eq_H_SU2_pure} around zero baryon density $b=0$ (where the baryon mass gap opens).

We observed that the presence of fermionic degrees of freedom affects only marginally the behavior of the gauge degrees of freedom of \cref{gauge_observables1}-\eqref{gauge_observables2}, albeit the magneto-electric transition becomes smoother at lower $m$ values (see also \cref{app_ED_results}).

For $m$ sufficiently large ($m{>}m^{*}(g)$), matter fields play a minor role (trivial phase). 
The Hamiltonian recovers the topological properties of the pure theory \cref{res_topology} (see also \cref{app_topology}) but no spontaneous topological order survives outside the magneto-electric transition $g^2{\sim} 2(1)$.

Conversely, for small masses $m{<} m^{*}(g)$, \cite{Engelhardt2000}, we observe an emergent color-density of the matter fields, only in the proximity of the magneto-electric transition. 
Such observation is compatible with the existence of a deconfined critical phase in the region where electric and magnetic fields are maximally frustrated (see also \cref{pure_fluctuations}).
Elsewhere, the system behaves like a gapless liquid of colorless baryons and anti-baryons.
The collective behavior towards the continuum limit is particularly intriguing, as it can exhibit both trivial or baryon superfluid phase depending on the quark ratio $\alpha_c$.

\section{Conclusions}
\label{sec_conclusions}
In this work, we analyzed, employing TN numerical simulations, a non-Abelian Yang-Mills LGTs in two spatial dimensions, with dynamical matter and hardcore gluons.
Our focus on this physical setting is motivated by the wide use of the latter as a paradigmatic model to address fundamental properties that could be relevant for high-dimensional QCD.
For instance, standard MC simulations have highlighted intriguing effects, such as the \emph{dimensional reduction} \cite{Ambjorn1984, Ambjorn1984a}, the compatibility with \emph{string theory} \cite{Ambjorn1984b, Ambjorn1984c}, and the possibility of accessing features of the \emph{continuum theory} already at small correlation lengths \cite{Berg1981}.

In summary, we have investigated in detail both the zero and finite baryon number density regimes, where MC methods are severely limited due to the sign problem.
Our results confirm TN methods as a reliable approach to addressing the non-perturbative phenomena of LGTs, capable of accessing strong coupling regimes as well as finite baryon number densities.

Despite the truncation of the gauge field, by exploiting numerical estimations of various observables, we inferred quite a few qualitative and quantitative observations concerning the zero-temperature phase diagram of the model.
First, when approaching the continuum limit ($a{\to}0$ at fixed $m_0$, $\alpha_c$) SU(2) baryons and anti-baryons become the actual quasiparticles of the theory. 
Interestingly, baryons seem to be able to condense into a superfluid phase for a sufficiently large quark ratio $\alpha_c{\geq}\alpha_c^{*}(m_0)$, that is, if their color charge is strong enough.

In the parameter regime at $g^{2}{\sim}2(1)$, where the electric term and the magnetic term are maximally frustrated, and electric and magnetic field fluctuations are proportional, we witnessed more exotic physics: at low quark masses, the system manifests colorful matter sites, possibly indicating a quark-deconfined regime, such as a quark-gluon plasma. 
At high quark masses, the system encounters a degeneracy between topological sectors (string symmetries in periodic boundary conditions), possibly signaling the emergence of a topological order reminiscent of the Toric code.

From a theoretical perspective, the studied Hamiltonian describes the interaction between flavorless 2-color fermionic matter and hardcore boson gauge fields (encoded as the $(0{\otimes} 0){\oplus}(\frac{1}{2}{\otimes}\frac{1}{2})$ representation).
Considering larger representations in the gauge Hilbert space (following the prescription detailed in \cref{app_model}) would be a natural extension of this work and an improved approximation of the continuous gauge field theory.

A larger truncation becomes substantial in the small coupling limit, where the Hamiltonian is dominated by the magnetic interaction, which is non-local and non-diagonal in the representation basis developed in \cref{app_model}. 
This makes the model significantly entangled and challenging to be numerically attacked via TNs.

As an outlook of this work, we plan to develop an analogous formalism in the magnetic basis, where plaquette terms are diagonal \cite{Kaplan2020, Haase2021, Paulson2021}.
This change of basis should ease TN simulations, which in our case are limited to finite system sizes, but anyway larger than the state-of-the-art of quantum-inspired or quantum simulations of non-Abelian LGTs \cite{Atas2021, Ciavarella2021, Ciavarella2023}. 
Accessing larger system sizes would be a substantial advantage, as it would enable the characterization of correlation functions not distorted by finite-size effects. Correspondingly, larger sizes would allow for studying magnetic effects at small coupling, as in MC simulations \cite{Kiskis1983, Kiskis1984, Hietanen2006}.

To overcome these limitations (finite gauge representation and finite system sizes), further developments of the numerical simulation architecture are also required: on the hardware side, the possibility of running the computation on a (pre)exascale HPC environment, while on the software side the development of new and improved TN-based algorithms. 
The latter will be achieved by exploiting the augmented TTN ansatz, which drastically enhances the capability of representing area law-states in high dimensions \cite{Felser2021}. 
These steps will be fundamental for the long-term goal of applying TN methods to large-scale lattice QCD in three spatial dimensions and ultimately address open, secular research problems, such as confinement and asymptotic freedom.

From an experimental viewpoint, the dressed-site formalism developed to build the Hamiltonian could be encoded on quantum hardware. In this perspective, the results and the methods presented in this work represent essential tools for benchmarking and validating current and future experimental implementations \cite{Meurice2022, DiMeglio2023, Zhang2023, Su2023}. 

\section{Acknowledgements}
We are thankful to Luca Tagliacozzo, Torsten V. Zache, Marcello Dalmonte, and Elisa Ercolessi for their enlightening suggestions concerning the topological characterization of the model. 
G.C. thanks Marco Rigobello for the precious discussions and the use of the \textsc{simsio} GitHub repository managing the I/O of simulations \cite{Rigobello2023}.
Authors acknowledge financial support to this research work
by the Italian Ministry of the University and Research (MUR) via PRIN2017, and PRIN2022 project TANQU;
by the European Union via QuantERA projects QuantHEP and T-NISQ, 
via Quantum Flagship project PASQuanS2,
and via NextGenerationEU (PNRR) project CN00000013 -  Italian Research Center on HPC, Big Data and Quantum Computing; 
by the WCRI-Quantum Computing and Simulation Center of Padova University,
by Fondazione CARIPARO, by Progetti Dipartimenti di Eccellenza via project Frontiere Quantistiche (FQ), and by the INFN project QUANTUM. G.M. is partially supported by UNIBA through the 2023-UNBACLE-0244025 grant and by INFN/NPQCD project.
We acknowledge computational resources by the Cloud Veneto, CINECA, the BwUniCluster, and the University of Padova Strategic Research Infrastructure Grant 2017: CAPRI: Calcolo ad Alte Prestazioni per la Ricerca e l'Innovazione.
The authors are also grateful to the Mainz Institute for Theoretical Physics (MITP) of the DFG Cluster of Excellence PRISMA* (Project ID 39083149) for its hospitality and partial support during the completion of this work. 

\appendix
\section{Dimensional analysis and continuum limit location}
\label{app_dimensional_analysis}
The simplest way to carry out dimensional analysis while locating the continuum limit in the space of coupling parameters is to consider the electric energy
\begin{equation}
    H_{\text{elec}} = \frac{\epsilon_c}{2} \int \mathcal{E}^2(\vb{x})  (d\vb{x})^D
\end{equation}
for a system of $D$ spatial dimensions, together with the Gauss' Law for electric fluxes
\begin{equation}
    \int \mathcal{E}(\vb{x}) \cdot \vb{u}_{\perp}
    (d\vb{x})^{D-1} = \frac{q_c}{\epsilon_c}.
\end{equation}
From these equations, it follows that, in dimensioned units (such as SI), the physical dimensions of the color-vacuum permittivity $\epsilon_c$ and the color-electric field $\mathcal{E}$ respectively read
\begin{equation}
\begin{aligned}
    \qty[\epsilon_c] &=(\text{charge})^2(\text{length})^{2-D}(\text{energy})^{-1}\\
    \qty[\mathcal{E}] &= (\text{charge})^{-1}(\text{length})^{-1}(\text{energy}).
    \end{aligned}
\end{equation}
To recast the problem onto a spatial lattice, we substitute
\begin{align}
 \int (d\vb{x})^D &\to  a^D \sum_{\vb{j},\vb*{\mu}} \,,&
 \mathcal{E}^2(\vb{x})&\to \frac{q^2_c a^{2-2D}}{\epsilon_c^2} E^2_{\vb{j},\vb*{\mu}},
\end{align}
where we introduced a lattice spacing $a$, a quark color-charge $q_c$, in such a way to obtain a dimensionless $E^2_{j,\mu}$ as in \cref{Casimir}. 
It is then possible to recast the charge in dimensionless units, precisely as
\begin{equation}
g = q_c \frac{a^{\frac{3-D}{2}}}{\sqrt{\hbar c \epsilon_c}},
\end{equation}
yielding the conversion
\begin{equation}
 H_{\text{elec}} =
 \frac{q^2_c a^{2-D}}{2 \epsilon_c} \sum_{\vb{j},\vb*{\mu}} E^2_{\vb{j},\vb*{\mu}}
  =
 g^2 \frac{c \hbar}{2 a} \sum_{\vb{j},\vb*{\mu}} E^2_{\vb{j},\vb*{\mu}}
\end{equation}
compatible with \cref{eq_H_SU2_pure}. 
Then, if we neglect quantum corrections to the scaling (see by contrast \cite{Creutz1980, Banuls2013}), it makes sense to assume that in the continuum $a \to 0$ limit the color-charge $q_c$ stays finite.
This assumption is perfectly reasonable for the SU(2) Yang-Mills theory in $D{=}2$ spatial dimensions, as it is known to be a \emph{super-renormalizable} theory \cite{Hamer1985}.
In this framework, $g^2$ has to scale linearly with $a$.
One can write:
\begin{align}
    g&= \sqrt{ a/a_g}, &
    \text{where}&& 
    a_g &= \hbar c \epsilon_c / q_c^2
\end{align}
is the (inverse square) color charge written as a length scale. 
Similarly, the bare mass can be expressed as:
\begin{equation}
 a_m = \frac{\hbar}{c m_0}.
\end{equation}
Correspondingly, moving toward the continuum limit $a{\to}0$, the electric energy coupling $g^2 c \hbar / 2a {=} c \hbar / 2 a_g$ and the mass coupling $m_{0} c^2 {=} \hbar c / a_m$ stay at the fixed ratio of
\begin{equation}
 \alpha_c = \frac{q_c^2}{ 2 m_0 c^2 \epsilon_c }
 = \frac{1}{2} \left(\frac{a_m}{a_g}\right),
\end{equation}
which is determined by the quark bare mass $m_0$ and its color charge $q_c$. 
As $\alpha_c$ is a dimensionless parameter not scaling with the lattice spacing, it plays a role equivalent to a \emph{fine-structure constant} (in two-spatial dimensions).

From a quantitative point of view, the continuum limit is reached when $a$ is the smallest length scale present in, or emergent from, the theory. 
Thus, first of all, we require that $a \ll a_m$ as well as $a \ll a_g$. 
Additionally, any emergent property, such as non-vanishing order parameters, must occur at wavelengths longer than $a$, basically $k \ll \frac{2 \pi}{a}$ (Infrared cutoff stability).

\section{Effective dressed-site model of truncated SU(2) Yang-Mills LGTs}
\label{app_model}
As discussed in \cref{model}, to make LGT Hamiltonians suitable for TN methods and quantum hardware, a finite-dimensional gauge-link Hilbert space is required. 
Here, we provide a comprehensive description of an effective truncated SU(2) Yang-Mills LGT that is valid for lattices of arbitrary spatial dimensions.

\begin{figure}
\centering
\includegraphics[width=1\columnwidth]{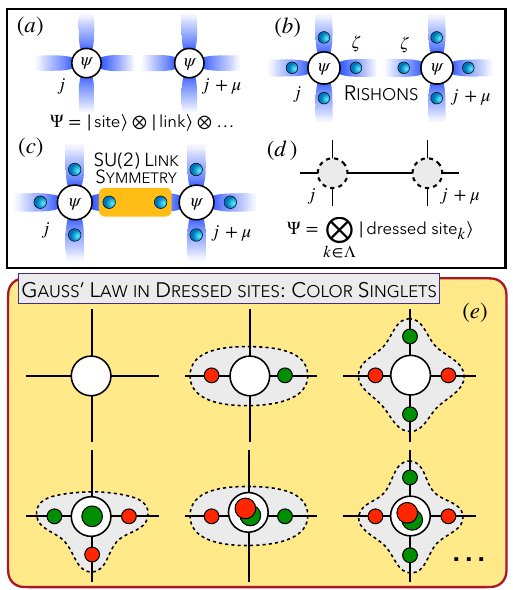}
\caption{Sketched representation of the approach developed in \cref{app_model}: 
{\it{(a)}} starting from the original formulation with matter sites and SU(2) gauge links, {\it{(b)}} we split the latter in pairs of rishon modes $\zeta$ defined in \cref{zeta_definition}, {\it{(c)}} constrain their dynamics with the SU(2) link symmetry in \cref{SU2_linksymmetry}, and {\it{(d)}} merge them with matter fields into dressed SU(2) gauge-singlets.}
\label{fig_dressed_site_scheme}
\end{figure}

On the trails of \cite{Tagliacozzo2013, Silvi2014}, we \emph{dress} every physical matter site with the information related to its adjacent gauge links. 
A pictorial scheme of this approach is outlined in \cref{fig_dressed_site_scheme}: starting from the original description matter fields on sites and gauge fields on links, {\it{(a)}} we truncate the SU(2) gauge group imposing an energy cut-off on the Casimir operator. 
Then, {\it{(b)}} we express each truncated gauge link as a pair of fermionic rishon mode $\zeta$ and {\it{(c)}} constrain their link dynamics according to the original SU(2) algebra. 
Ultimately, {\it{(d)}} we merge each of these modes to its adjacent matter site, ending up in a compact \emph{dressed-site} formalism.
The resulting effective Hamiltonian is made out of only bosonic operators and directly acts on the SU(2) gauge invariant Hilbert sub-space. 

Such an approach is general and valid for all the possible incremental truncations of the SU(2) gauge links.
Moreover, in the limit of an infinitely large spin irreducible representation of the SU(2) gauge group, it recovers all the properties of the original SU(2) Yang-Mills LGT. 
Nonetheless, the use of this approach in the minimal truncation of SU(2) has been used to achieve non-trivial results as the ones discussed in \cref{results}.

\subsection{Truncating the SU(2) gauge group}
\label{sub_app_SU2_QLM}
Let us start recalling the properties of the original SU(2) LGT.
In the presence of single-flavor matter fields with SU(2)-color 1/2 $\qty{\red{r},\green{g}}$, expressed in terms of Dirac fermions $\psi_{\vb{j}, \alpha}$  and located at the lattice sites $\vb{j}\in \Lambda$, gauge fields $\hat{E}_{\vb{j},\vb{j}+\vb*{\mu}}$ and $\hat{U}_{\vb{j},\vb{j}+\vb*{\mu}}^{\alpha,\beta}$ live on lattice links $(\vb{j},\vb{j}+\vb*{\mu})$ and generate the SU(2) gauge algebra in \cref{Gaugealgebra}.

To truncate the continuous SU(2) gauge group, we express it in terms of the irreducible representation (irrep) basis \cite{Zohar2015, Burgio2000}.
As SU(2) admits a \emph{quasi-real}-representation, where the fundamental and the anti-fundamental representations coincide, $\forall \vb{j},\vb*{\mu} {\in} \Lambda$, the gauge Hilbert space of the $(\vb{j},\vb{j}+\vb*{\mu})$ link can be written as:
\begin{equation}
\begin{aligned}
    \mathcal{H}_{\text{link}}&{=}\qty{\ket{j,m_{L},m_{R}}},&
    \text{where}&&
-j{\leq} m_{L(R)}{\leq} j
\end{aligned}
\label{eq_gauge_hilbert_space}
\end{equation}
is the corresponding third spin component associated with the left (right) side of the SU(2) link-irrep $j$.
Correspondingly, the parallel transporter $\hat{U}_{\vb{j},\vb{j}+\vb*{\mu}}^{\alpha\beta}$ act as follows:
\begin{equation}
\bra{j'm'_{L}m'_{R}}\hat{U}^{\alpha \beta}\ket{jm_{L}m_{R}}{=}
 \overline{C^{j m_{L}}_{\frac{1}{2},\alpha; j'm'_{L}} }
 C^{j' m'_{R}}_{j m_{R}; \frac{1}{2},\beta}
\label{eq_U_property1}
\end{equation}
where the $C$ symbols are the Clebsh-Gordan coefficients for SU(2) \cite{Zohar2015}.
In hopping terms, the action of $\hat{U}$ has to match the fundamental irrep of the matter field, whose Hilbert space in the Fock space, can be written as:
\begin{equation}
    \mathcal{H}_{\text{site}}=\qty{
    \ket{\Omega},
    \psi^{\dagger}_{\red{r}}\ket{\Omega},
    \psi^{\dagger}_{\green{g}}\ket{\Omega},
    \psi^{\dagger}_{\red{r}}
    \psi^{\dagger}_{\green{g}}\ket{\Omega}},
    \label{eq_matter_hilbert_space}
\end{equation} 
whose action in the irrep basis $\ket{j;m}$ corresponds to the following SU(2) charges:
\begin{equation}
\begin{split}
    \mathcal{H}_{\text{site}}=
    \qty{\ket{0;0},\ket{\mbox{$\frac{1}{2}$};\mbox{$\frac{1}{2}$}},\ket{\mbox{$\frac{1}{2}$};-\mbox{$\frac{1}{2}$}},\ket{0;0}}.
\end{split}
\end{equation}
In this basis, the electric field $\hat{E}_{\vb{j},\vb{j}+\vb*{\mu}}$ is diagonal, and
\begin{equation}
\bra{j'm'_{L}m'_{R}}\hat{E}^{2}\ket{jm_{L}m_{R}}{=}C_{2}(j)\delta_{j,j'}\delta_{m_{L}, m_{L}'}\delta_{m_{R},m_{R}'},
\end{equation}
where $C_{2}(j) = j(j+1)$ is the quadratic Casimir.

Then, the truncation of the gauge fields is applied by imposing a cutoff $\Theta$ on the $\hat{E}^{2}$ spectrum, keeping the irreps $j\leq j_{\text{max}}$  such that $C_{2}(j_{\text{max}})\leq\Theta$.
Such a truncation preserves the SU(2) gauge algebra in \cref{Gaugealgebra}. Namely, truncated $\hat{U}^{\alpha\beta}$ fields satisfy the following left and right gauge transformations:
\begin{equation}
\begin{aligned}
  [\hat{\vb{L}}_{j_{\text{max}}},\hat{U}_{\alpha\beta}]&=-\sum_{\gamma}\frac{\vb*{\sigma}_{\alpha\gamma}}{2}\hat{U}_{\gamma\beta}\\
  [\hat{\vb{R}}_{j_{\text{max}}},\hat{U}_{\alpha\beta}]&=+\sum_{\gamma}\hat{U}_{\alpha\gamma}\frac{\vb*{\sigma}_{\gamma \beta}}{2},
    \label{LR_generators}
\end{aligned}
\end{equation}
where $\hat{\vb{L}}_{j_{\text{max}}}$ and $\hat{\vb{R}}_{j_{\text{max}}}$ are (truncated) generators of the left- and right-handed groups of SU(2) transformations. They can be expressed as the block-diagonal direct sum of spin matrices $\vb{S}_{j}$ in consecutive $j$-representations from the smallest ($j=0$) to the largest one ($j=j_{\text{max}}$):
\begin{equation}
\begin{aligned}
\hat{\vb{L}}_{j_{\text{max}}}&{=}\bigoplus_{j=0}^{j_{\text{max}}}\qty(\vb{S}_{j}{\otimes} \mathbb{1}_{j}){=}\text{diag}(\vb{S}_{0}{\otimes}\mathbb{1}_{0},\dots, \vb{S}_{j_{\text{max}}}{\otimes} \mathbb{1}_{j_{\text{max}}})\\
\hat{\vb{R}}_{j_{\text{max}}}&{=}\bigoplus_{j=0}^{j_{\text{max}}}\qty(\mathbb{1}_{j}{\otimes}\vb{S}_{j}){=}\text{diag}(\mathbb{1}_{0}{\otimes}\vb{S}_{0},\dots,\mathbb{1}_{j_{\text{max}}}{\otimes}\vb{S}_{j_{\text{max}}} ),
\end{aligned}
\label{eq_LR_generator}
\end{equation}
in such a way that, $\forall m,n\in\qty{x,y,z}$, 
$\qty[L^{m}_{j_{\text{max}}},R^{n}_{j_{\text{max}}}]=0$.
The truncation keeps $\hat{U}$ unitary only as long as it acts on spin shells with $j<j_{\text{max}}$. 
Correspondingly, Wilson loops stay unitary as long as the outer spin shell $j=j_{\text{max}}$ is nowhere populated. 
Namely, it implies that, $\forall j < j_{\text{max}}$
\begin{equation}
\begin{aligned}
       {\sum_{\beta\gamma}}{U_{\alpha\beta}}{\ket{0{;}0{,}0}}{\otimes} {U_{\beta\gamma}}{\ket{j{;}m_{L}{,}m_{R}}}{\otimes} {U_{\gamma\delta}}{\ket{0{;}0{,}0}}
\end{aligned}
\end{equation}
displays the same norm of $U_{\alpha\delta}\ket{000}$ $\forall \alpha, \delta$, that is $1/\sqrt{2}$.
Ultimately, we require the parallel transporter to display \emph{spatial reflection symmetry}. Namely:
\begin{equation}
    \hat{U}^{\dagger}_{\alpha\beta}=-\hat{\mathcal{F}}\hat{U}_{\beta\alpha}\hat{\mathcal{F}},
\end{equation}
where $\hat{\mathcal{F}}$ is the \emph{swap operator} on a gauge link:
\begin{align}
\hat{\mathcal{F}}{=}\sum_{j,m_{L},m_{R}}{\ket{j; m_{L},m_{R}}}{\bra{j;m_{R},m_{L}}}{=}\hat{\mathcal{F}}^{-1}{=}\hat{\mathcal{F}}^{\dagger}
\end{align}
With all these premises, the resulting truncated SU(2) Yang-Mills Hamiltonian of \cref{eq_H_SU2_full}-\eqref{eq_H_SU2_pure} will act on quantum-many-body (QMB) states like the following:
\begin{equation}
  \ket{\Psi}=\bigotimes_{\vb{j}\in \Lambda}\bigotimes_{\vb*{\mu}}\ket{\text{site}}_{\vb{j}}\otimes\ket{\text{link}}_{\vb{j},\vb{j}+\vb*{\mu}},
  \label{Psi-QMB}
\end{equation}
where gauge-link and matter sites degrees of freedom live respectively in \cref{eq_gauge_hilbert_space} and \cref{eq_matter_hilbert_space}.
Notice that matter and gauge fields display different statistics. 
In these terms, we expect $U_{\vb{j},\vb{j}+\vb*{\mu}}^{\alpha\beta}$ to be \emph{mutually bosonic}, as it commutes with all the matter-fields operators:
\begin{equation}
  \qty[\hat{U}_{\vb{j},\vb{j}+\vb*{\mu}}^{\alpha\beta},\hat{\psi}^{(\dagger)}_{\vb{j},\alpha}]=0
  \qquad \forall \vb{j},\forall \vb*{\mu}, \forall \alpha,\beta
\end{equation}
and \emph{purely local}, as its link-algebra commutes with the one of any other link:
\begin{align}
  \qty[\hat{U}_{\vb{j},\vb{j}+\vb*{\mu}}^{\alpha\beta},\hat{U}_{\vb{j'},\vb{j'}+\mu'}^{\gamma\delta}]&{=}0 &
  \forall &\vb{j}\neq \vb{j'},\vb*{\mu} \neq \vb*{\mu'}, \forall \alpha,\beta,\gamma,\delta
\end{align}
Therefore, any numerical simulation of such an LGT has to consider both fermionic and bosonic anti-commutation rules. Moreover, among all the possible QMB states in \cref{Psi-QMB}, we must select only the ones where the SU(2)-Gauss law is locally satisfied (see the \emph{(e)} panel of \cref{fig_dressed_site_scheme}). 

\subsection{SU(2) fermionic rishon modes for arbitrary representations}
\label{sub_app_rishon_formalism}
It is, of course, possible to generalize the rishon decomposition of $\hat{U}^{\alpha\beta}$ to arbitrary truncation of the maximum allowed spin shell $j_{\text{max}}$, although at a (manageable) added cost.
Starting from a given spin shell $j$, we have to separately account for the action when both rishons are increased to shell $j+\frac{1}{2}$, and both are decreased to shell $j-\frac{1}{2}$. 
We can then decompose $\hat{U}^{\alpha\beta}$ as follows:
\begin{equation}
 \hat{U}^{\alpha\beta}_{\vb{j}, \vb{j}+\vb*{\mu}}{=}\hat{\zeta}_{A,\vb{j},\vb*{\mu}}^{\alpha} \hat{\zeta}_{B,\vb{j}+\vb*{\mu},-\vb*{\mu}}^{\beta\dagger}+\hat{\zeta}_{B,\vb{j},\vb*{\mu}}^{\alpha} \hat{\zeta}_{A,\vb{j}+\vb*{\mu},-\vb*{\mu}}^{\beta\dagger},
 \label{eq_SU2_U_definition}
\end{equation}
where the two $\zeta$-rishon species, A and B, act respectively as raising and lowering the spin shell of the SU(2) gauge irrep. Interestingly, they are related to each other as:
\begin{align}
    \hat{\zeta}_{A}^{\alpha} &= i \sigma^{y}_{\alpha,\beta} \hat{\zeta}_{B}^{\beta\dagger} &
    \hat{\zeta}_{A}^ {\alpha\dagger} = i \sigma^{y}_{\alpha,\beta} \hat{\zeta}_{B}^{\beta}
\end{align}
We can then rewrite \cref{eq_SU2_U_definition} just in terms of one species, e.g. B. 
Dropping the index, i.e. $\hat{\zeta}_{B}^{\alpha}{=}\hat{\zeta}^{\alpha}$, it holds:
\begin{equation}
 \hat{U}^{\alpha\beta}_{\vb{j}, \vb{j}+\vb*{\mu}}{=} 
 i \sigma^{y}_{\alpha\gamma} \hat{\zeta}_{\vb{j},\vb*{\mu}}^{\gamma\dagger} \hat{\zeta}_{\vb{j}+\vb*{\mu},-\vb*{\mu}}^{\beta\dagger} 
 + i \sigma^{y}_{\beta\gamma} \hat{\zeta}_{\vb{j},\vb*{\mu}}^{\alpha} \hat{\zeta}_{\vb{j}+\vb*{\mu},-\vb*{\mu}}^{\gamma}
\end{equation}
or equivalently
\begin{equation}
 \begin{aligned}
&\hat{U}^{\red{r}\red{r}}_{\vb{j}, \vb{j}+\vb*{\mu}}=\hat{\zeta}_{\vb{j},\vb{j}+\vb*{\mu}}^{\green{g}\dagger} \hat{\zeta}_{\vb{j}+\vb*{\mu},-\vb*{\mu}}^{\red{r}\dagger} +
\hat{\zeta}_{\vb{j},\vb{j}+\vb*{\mu}}^{\red{r}} \hat{\zeta}_{\vb{j}+\vb*{\mu},-\vb*{\mu}}^{\green{g}}\\
&\hat{U}^{\red{r}\green{g}}_{\vb{j}, \vb{j}+\vb*{\mu}} =\hat{\zeta}_{\vb{j},\vb{j}+\vb*{\mu}}^{\green{g}\dagger} \hat{\zeta}_{\vb{j}+\vb*{\mu},-\vb*{\mu}}^{\green{g}\dagger} -
\hat{\zeta}_{\vb{j},\vb{j}+\vb*{\mu}}^{\red{r}} \hat{\zeta}_{\vb{j}+\vb*{\mu},-\vb*{\mu}}^{\red{r}}\\
&\hat{U}^{\green{g} \red{r}}_{\vb{j}, \vb{j}+\vb*{\mu}}=-\hat{\zeta}_{\vb{j},\vb{j}+\vb*{\mu}}^{\red{r}\dagger} \hat{\zeta}_{\vb{j}+\vb*{\mu},-\vb*{\mu}}^{\red{r}\dagger} + \hat{\zeta}_{\vb{j},\vb{j}+\vb*{\mu}}^{\green{g}} \hat{\zeta}_{\vb{j}+\vb*{\mu},-\vb*{\mu}}^{\green{g}}\\
&\hat{U}^{\green{g} \green{g}}_{\vb{j}, \vb{j}+\vb*{\mu}}= -\hat{\zeta}_{\vb{j},\vb{j}+\vb*{\mu}}^{\red{r}\dagger} \hat{\zeta}_{\vb{j}+\vb*{\mu},-\vb*{\mu}}^{\green{g}\dagger}- \hat{\zeta}_{\vb{j},\vb{j}+\vb*{\mu}}^{\green{g}} \hat{\zeta}_{\vb{j}+\vb*{\mu},-\vb*{\mu}}^{\red{r}}
 \end{aligned}
\end{equation}
For a chosen truncation $j_{\text{max}}$ of the SU(2) irreducible representation, $\zeta$-rishons are defined as follows:
\begin{equation}
 \hat{\zeta}^{\green{g}(\red{r})}{=}\qty[\sum_{j=0}^{j_{\text{max}}-\frac{1}{2}}\sum_{m=-j}^{j}\chi(j,m,\green{g}(\red{r})) \ket{j,m}\bra{j{+}\mbox{$\frac{1}{2}$},m{\cmp}\mbox{$\frac{1}{2}$}}]_F
 \label{eq_SU2_general_rishon}
\end{equation}
where the function $\chi(j,m,\alpha)$ reads
\begin{equation}
    \chi\qty(j,m,\green{g}(\red{r}))=\sqrt{\frac{j\cmp m+1}{\sqrt{(2j+1)(2j+2)}}}.
\end{equation}
It is possible to show that this construction is indeed compatible with the explicit form of the parallel transport reported in \cref{eq_U_property1}.

\subsection{SU(2) Rishon Parity}
\label{sec_SU2_rishon_parity}
By construction, $\zeta$-rishons anti-commute among themselves at different orbitals and with matter fields:
\begin{align}
  \qty{\hat{\zeta}_{\vb{j},\vb*{\mu}}^{\alpha},\hat{\zeta}_{\vb{j}+\vb*{\mu},-\vb*{\mu}}^{\beta}}&=0
  &
  \qty{\hat{\zeta}_{\vb{j},\vb*{\mu}}^{\alpha},\hat{\psi}_{\vb{j},\beta}}&=0& \forall \alpha,\beta
  \label{eq_SU2_zeta_commmutations}
\end{align}
To satisfy \cref{eq_SU2_zeta_commmutations}, we need to characterize them as fermion operators properly.  
For a fermionic Quantum Many-Body system with particles arbitrarily sorted along a certain path, any tensor product of fermionic operators should take into consideration the proper anti-commutation rules. 
Namely, a generic fermionic operator $\hat{F}_{\vb{j}}$ acting on the $\vb{j}^{th}$ position along the path reads:
\begin{equation}
    \hat{F}_{\vb{j}}=\dots P_{\vb{j-2}}\otimes P_{\vb{j-1}}\otimes F_{\vb{j}} \otimes \mathbb{1}_{\vb{j+1}}\otimes \mathbb{1}_{\vb{j+2}}\dots 
    \label{fermionic_qmb_op}
\end{equation}
where $P_{\vb{j}}=P_{\vb{j}}^{\dagger}=P_{\vb{j}}^{-1}$ is a fermion parity operator that gets inverted after the action of a fermionic operator:
\begin{equation}
\qty{P_{\vb{j}},F_{\vb{j}}}=0 \qquad 
\qty[P_{\vb{j}},F_{\vb{j'}\neq \vb{j}}]=0 
\qquad \forall \vb{j},\vb{j'}\in \Lambda
\label{fermion_parity_commutation}
\end{equation}
Therefore, matter fields admit their notion of parity satisfying \cref{fermion_parity_commutation}. For Dirac fermions, we have:
\begin{align}
  \hat{\psi}_{\text{Dirac}} &= \qty( 
  \begin{array}{cc}
   0 & 1 \\
   0 & 0
   \end{array})_F&
   \hat{P}_{\text{Dirac}} &= \qty( 
  \begin{array}{cc}
   +1 & 0 \\
   0 & -1
   \end{array})  
\end{align}
where the subscript $F$ is a reminder that the $\hat{\psi}$ matrix is meant 'as a fermion', with the global action in \cref{fermionic_qmb_op}.
Similarly, as for Majorana fermions, we have:
\begin{align}
    \hat{\gamma}_{\text{Majorana}} &{=} \qty( 
  \begin{array}{cc}
   0 & 1 \\
   1 & 0
   \end{array})_F&
   \hat{P}_{\text{Majorana}} &{=} \qty( 
  \begin{array}{cc}
   {+}1 & 0 \\
   0 & {-}1
   \end{array})  
\end{align}
Being fermions, also $\zeta$-rishons satisfy \cref{eq_SU2_zeta_commmutations}. 
We define the SU(2) rishon parity operator $P_{\zeta}$ with an even ($+1$) parity sector on \emph{integer} irreps and odd ($-1$) sector on \emph{semi-integer} ones. 

Correspondingly, the parallel transporter $\hat{U}_{\vb{j},\vb{j}+\vb*{\mu}}^{\alpha\beta}$ reads:
\begin{equation}
  \begin{split}
    \hat{U}_{\vb{j},\vb{j}+\vb*{\mu}}^{\alpha\beta}{=}& 
 i\sigma^{y}_{\alpha\gamma} \hat{\zeta}_{\vb{j},\vb*{\mu}}^{\gamma\dagger} \hat{\zeta}_{\vb{j}+\vb*{\mu},-\vb*{\mu}}^{\beta\dagger} 
 + i\sigma^{y}_{\beta\gamma} \hat{\zeta}_{\vb{j},\vb*{\mu}}^{\alpha} \hat{\zeta}_{\vb{j}+\vb*{\mu},-\vb*{\mu}}^{\gamma}\\
    {=}&+i\sigma^{y}_{\alpha\gamma}\qty(\zeta_{\vb{j},\vb*{\mu}}^{\gamma\dagger}\cdot P_{\zeta, \vb{j},\vb*{\mu}})\otimes \zeta_{\vb{j}+\vb*{\mu},-\vb*{\mu}}^{\beta\dagger}\\
    &+i\sigma^{y}_{\beta\gamma}\qty(\zeta_{\vb{j},\vb*{\mu}}^{\alpha}\cdot P_{\zeta, \vb{j},\vb*{\mu}})\otimes \zeta_{\vb{j}+\vb*{\mu},-\vb*{\mu}}^{\gamma}
  \end{split}
\end{equation}

\subsection{SU(2) rishon algebra}
\label{sec_SU2_rishon_algebra}
Instead of relying on two separate SU(2) generators, $\vb{L}$ and $\vb{R}$, $\zeta$-rishons have a unique gauge transformation algebra.  
The generator of SU(2) gauge rotations upon the $\zeta$-rishon space reads:
\begin{equation}
\hat{\vb{T}}_{j_{\text{max}}}=\bigoplus_{j=0}^{j_{\text{max}}}\vb{S}_{j}=\text{diag}(\vb{S}_{0},\vb{S}_{1},\dots \vb{S}_{j_{\text{max}}})
\label{eq_T_generator}
\end{equation}
By construction, $\zeta$ operators are SU(2) covariant, as they transform as follows:
\begin{equation}
\begin{aligned}
  \qty[\hat{\zeta}^{\alpha},\hat{\vb{T}}]&{=}\frac{1}{2}\sum_{\beta}\vb*{\sigma}_{\alpha\beta}\hat{\zeta}^{\beta}&
  \qty[\hat{\vb{T}},\hat{\zeta}_{\alpha}^{\dagger}]&{=}\frac{1}{2}\sum_{\beta}\hat{\zeta}^{\beta\dagger}\vb*{\sigma}_{\beta\alpha}
\end{aligned}
\label{zeta_algebra}
\end{equation}
Moreover, $\hat{\vb{T}}$ is genuinely \emph{local}, as for $\forall \vb{j}\neq\vb{j'}\, \forall \vb*{\mu}\neq\vb*{\mu}'$:
\begin{align}
\big[\hat{\vb{T}}_{\vb{j}+\vb*{\mu}},\hat{\zeta}_{\vb{j'}+\vb*{\mu}'}^{\alpha}\big]&{=}\big[\hat{\vb{T}}_{\vb{j}+\vb*{\mu}},\hat{\psi}_{\vb{j'}, \alpha}\big]{=}0&\forall \alpha\in\qty{\red{r},\green{g}}&
\end{align}
   
We can easily recover the left- and right-handed sides generators of the gauge field at link $(\vb{j},\vb{j}+\vb*{\mu})$ as:
\begin{equation}
\begin{aligned}
  \hat{\vb{L}}_{\vb{j},+\vb*{\mu}}&{=}\hat{\vb{T}}_{\vb{j}{,}{+}\vb*{\mu}}{\otimes} \mathbb{1}_{\vb{j}+\vb*{\mu},-\vb*{\mu}}&
  \hat{\vb{R}}_{\vb{j},+\vb*{\mu}}&{=}\mathbb{1}_{\vb{j},+\vb*{\mu}}{\otimes}\hat{\vb{T}}_{\vb{j}+\vb*{\mu},-\vb*{\mu}}
\end{aligned}
\label{eq_LR_rishon_generator}
\end{equation}
Since in the SU(2) group the fundamental and the anti-fundamental representations coincide, the rishon formalism is meaningful as long as the quadratic Casimir operator of the two sides of the link coincide as in \cref{Casimir}: 
\begin{equation}
  \abs{\hat{\vb{L}}_{\vb{j}, +\vb*{\mu}}}^{2}=\abs{\hat{\vb{R}}_{\vb{j}+\vb*{\mu}, -\vb*{\mu}}}^{2}
  \label{SU2_linksymmetry}
\end{equation}
Thanks to \cref{SU2_linksymmetry}, the two rishons of the link are in the same SU(2) irrep, and the parallel transport in \cref{eq_SU2_U_definition} coincides with the one in \cref{eq_U_property1}. 
Correspondingly, the Casimir operator of the $(\vb{j},\vb{j}+\vb*{\mu})$ link in \cref{Casimir} can be expressed as:
\begin{equation}
  \begin{split}
  &\hat{E}^{2}_{\vb{j},\vb{j}+\vb*{\mu}}
  = \frac{1}{2}\qty[\abs{\hat{\vb{L}}_{\vb{j},+\vb*{\mu}}}^2 +\abs{\hat{\vb{R}}_{\vb{j}+\vb*{\mu},-\vb*{\mu}}}^2]
  \end{split}
  \label{eq_Electric_casimir}
\end{equation}
which looks explicitly symmetric under link reversal.

\subsection{Example: minimally truncated SU(2) gauge link}
\label{sec_SU2_example}
As an example, we consider the smallest non-trivial representation of the gauge fields, obtained truncating the Casimir up to $j_{\text{max}}=\frac{1}{2}$ \cite{Horn1981, Orland1990, Yao2023, Muller2023}. 
This truncation is the one adopted in \cref{results} and corresponds to the following 5-dimensional gauge-link Hilbert space:
\begin{equation}
  \mathcal{H}_{\text{link}}=
  \qty{\ket{0,0},\ket{\red{r},\red{r}},\ket{\red{r},\green{g}},
  \ket{\green{g},\red{r}},\ket{\green{g},\green{g}}},
  \label{link_5D_Hilbert_space}
\end{equation}
Within this representation, we can then define the corresponding versions of the truncated gauge fields. 
As for the parallel transport, we have \cite{Zohar2015}:
\begin{equation}
    {U_{\alpha\beta}}{=}{\frac{1}{\sqrt{2}}}
    \qty(
      \begin{array}{@{\hspace{0.1ex}}c@{\hspace{0.1ex}}|@{\hspace{0.1ex}}c@{\hspace{0.2ex}}c@{\hspace{0.2ex}}c@{\hspace{0.2ex}}c@{\hspace{0.1ex}}}
    0&{+}\delta_{\alpha\red{r}}\delta_{\beta\green{g}} 
     &{-}\delta_{\alpha\red{r}}\delta_{\beta\red{r}}  
     &{+}\delta_{\alpha\green{g}}\delta_{\beta\green{g}}
     &{-}\delta_{\alpha\green{g}}\delta_{\beta\red{r}} \\
    \hline
    {-}\delta_{\alpha\green{g}} \delta_{\beta\red{r}}&0&0&0&0\\
    {-}\delta_{\alpha\green{g}} \delta_{\beta\green{g}}&0&0&0&0\\
    {+}\delta_{\alpha\red{r}} \delta_{\beta\red{r}}&0&0&0&0\\
    {+}\delta_{\alpha\red{r}} \delta_{\beta\green{g}}&0&0&0&0\\
  \end{array})
    \label{parallel_transport}
\end{equation}
where the $1/\sqrt{2}$ factor ensures that the hopping term preserves the state norm on its support. Correspondingly, the quadratic Casimir operator in \cref{Casimir}:
\begin{equation}
  E^{2}=\frac{3}{4}
  \qty(\begin{array}{c|cccc}
      0&0&0&0&0\\
      \hline
      0&1&0&0&0\\
      0&0&1&0&0\\
      0&0&0&1&0\\
      0&0&0&0&1\\
  \end{array})
  \label{Casimir2}
\end{equation}
Correspondingly, $\zeta$-rishons in \cref{eq_SU2_general_rishon} reduces to:
\begin{align}
  \hat{\zeta}_{\red{r}} &= \frac{1}{\sqrt[4]{2}} \qty( 
  \begin{array}{c|cc}
   0 & 1 & 0 \\
   \hline
   0 & 0 & 0 \\
   0 & 0 & 0 \\  
   \end{array})_{F}&
  \hat{\zeta}_{\green{g}} &= \frac{1}{\sqrt[4]{2}} \qty( 
  \begin{array}{c|cc}
   0 & 0 & 1 \\
   \hline
   0 & 0 & 0 \\
   0 & 0 & 0 \\  
   \end{array})_{F}
   \label{zeta_definition}
 \end{align}
 with the corresponding parity operator:
 \begin{equation}
  \hat{P}_{\zeta} = \qty( 
  \begin{array}{c|cc}
   1 & 0 & 0 \\
   \hline
   0 & -1 & 0 \\
   0 & 0 & -1 \\  
   \end{array})
   \label{eq_SU2_rishon_parity}
\end{equation}
and SU(2) rishon-generators
\small
\begin{equation}
    \begin{aligned}
  \hat{T}_{1/2}^{x}&{=}{\frac{1}{2}}
  \qty(\begin{array}{@{\hspace{0.1ex}}c@{\hspace{0.3ex}}|c@{\hspace{0.3ex}}c@{\hspace{0.1ex}}}
      0& \\
      \hline
      & 0&1\\
      & 1&0\\
  \end{array})&
  \hat{T}_{1/2}^{y}&{=}{\frac{1}{2}}
  \qty(\begin{array}{@{\hspace{0.1ex}}c@{\hspace{0.3ex}}|c@{\hspace{0.3ex}}c@{\hspace{0.1ex}}}
      0& \\
      \hline
      & 0&{-}i\\
      & i&0\\
  \end{array})&
  \hat{T}_{1/2}^{z}&{=}{\frac{1}{2}}
  \qty(\begin{array}{@{\hspace{0.1ex}}c@{\hspace{0.3ex}}|c@{\hspace{0.3ex}}c@{\hspace{0.1ex}}}
      0& \\
      \hline
      & 1&0\\
      & 0&{-}1\\
  \end{array})
  \label{eq_T1/2}
\end{aligned}
\end{equation}
\normalsize
By definition, the spin-Hilbert space of every side of the link $(\vb{j},\vb{j}+\vb*{\mu})$  hosting a rishon mode is 3-dimensional:
\begin{equation}
  \mathcal{H}_{\vb{j},+\vb*{\mu}}=\qty{\ket{0},\ket{\red{r}},\ket{\green{g}}}=\mathcal{H}_{\vb{j}+\vb*{\mu}, -\vb*{\mu}}
\end{equation}
Correspondingly, the Hilbert space of the whole link $\mathcal{H}_{\text{link}}=\mathcal{H}_{\vb{j},+\vb*{\mu}}\otimes \mathcal{H}_{\vb{j}+\vb*{\mu}, -\vb*{\mu}}$ has 9 states.
To recover the original 5-dimensional space in \cref{link_5D_Hilbert_space}, we must impose the SU(2) link constraint defined in \cref{SU2_linksymmetry}.

\subsection{Constructing dressed site operators}
We have then all the ingredients to build a \emph{dressed-site} compact representation. In the case of a 2D lattice, one possible pictorial description of \emph{dressed-site} states reads:
\begin{align}
    \ket{\begin{array}{@{\hspace{0.1ex}}c@{\hspace{0.4ex}}c@{\hspace{0.4ex}}c@{\hspace{0.1ex}}}
        &\hat{\zeta}_{\vb{j},+\vb*{\mu}_{y}}&\\
        \hat{\zeta}_{\vb{j},-\vb*{\mu}_{x}}&\qty(\hat{\psi}_{\vb{j},\red{r}}\hat{\psi}_{\vb{j},\green{g}})&\hat{\zeta}_{\vb{j},+\vb*{\mu}_{x}}\\
        &\hat{\zeta}_{\vb{j},-\vb*{\mu}_{y}}&\\
    \end{array}},&&\text{where}&&
    \ket{\begin{array}{@{\hspace{0.1ex}}c@{\hspace{0.4ex}}c@{\hspace{0.4ex}}c@{\hspace{0.1ex}}}
        &5&\\
        2&(0,1)&4\\
        &3&\\
    \end{array}}
    \label{eq_SU2_dressed_site}
\end{align}
is a possible internal ordering to be used as in \cref{fermionic_qmb_op} when constructing composite operators out of matter fields and rishons inside the dressed site. 

We can start rewriting the SU(2) Yang-Mills Hamiltonian terms in \cref{eq_H_SU2_full}-\eqref{eq_H_SU2_pure} in terms of rishon modes.

\paragraph{Arrival operators}
Let us start with the hopping Hamiltonian term. Discarding all the pre-factors, we can focus on:
\begin{equation*}
    \begin{split}
\hat{h}&^{\text{hopping}}_{\vb{j},\vb{j}+\vb*{\mu}}=\sum_{\alpha,\beta}\hat{\psi}^{\dagger}_{\vb{j} ,\alpha} \hat{U}_{\vb{j},\vb{j}+\vb*{\mu}}^{\alpha\beta} \hat{\psi}_{\vb{j}+\vb*{\mu},\beta}\\
{=}&\sum_{\alpha,\beta}\hat{\psi}^{\dagger}_{\vb{j},\alpha}
\qty[\hat{\zeta}_{A,\vb{j},\vb*{\mu}}^{\alpha} \hat{\zeta}_{B,\vb{j}+\vb*{\mu},-\vb*{\mu}}^{\beta\dagger} 
 {+}\hat{\zeta}_{B,\vb{j},\vb*{\mu}}^{\alpha} \hat{\zeta}_{A,\vb{j}+\vb*{\mu},-\vb*{\mu}}^{\beta\dagger}]
 \hat{\psi}_{\vb{j}+\vb*{\mu},\beta}\\
 {=}&\qty[\hat{Q}^{\dagger}_{A,\vb{j},\vb*{\mu}}\hat{Q}_{B,\vb{j}+\vb*{\mu},-\vb*{\mu}}+\hat{Q}^{\dagger}_{B,\vb{j}}\hat{Q}_{A,\vb{j}+\vb*{\mu},-\vb*{\mu}}],
    \end{split}
\end{equation*}
where we defined two species of \emph{arrival} operators:
\begin{align}
\hat{Q}^{\dagger}_{A,\vb{j},\vb*{\mu}}&{=}\sum_{\alpha}\hat{\psi}^{\dagger}_{\vb{j},\alpha}\hat{\zeta}_{A,\vb{j},\vb*{\mu}}^{\alpha}&
\hat{Q}^{\dagger}_{B,\vb{j},\vb*{\mu}}&{=}\sum_{\alpha}\hat{\psi}^{\dagger}_{\vb{j},\alpha} \hat{\zeta}_{B,\vb{j},\vb*{\mu}}^{\alpha}
\label{eq_arrival_operators}
\end{align}
These operators' practical construction must be consistent with the internal ordering in \cref{eq_SU2_dressed_site}.

\paragraph{Matter number density operators}
Matter density operators acting on the dressed-site basis read:
\begin{equation}
    \begin{aligned}
        \hat{N}_{\vb{j},\red{r}}&
        =\psi^{\dagger}_{\vb{j}}\psi_{\vb{j},\red{r}}
        \otimes\mathbb{1}_{\vb{j},\green{g}}\bigotimes_{\vb*{\mu}} \mathbb{1}_{\vb{j},\vb*{\mu}}\\
        \hat{N}_{\vb{j},\green{g}}&
        =\mathbb{1}_{\vb{j},\red{r}}
    \otimes \psi^{\dagger}_{\vb{j}}\psi_{\vb{j},\green{g}}\bigotimes_{\vb*{\mu}} \mathbb{1}_{\vb{j},\vb*{\mu}}\\
    \end{aligned}
    \label{eq_number_operators1}
\end{equation}
They also give access to other observables like:
\begin{equation}
\begin{aligned}
    \hat{N}_{\vb{j},\text{tot}}&
    =\hat{N}_{\vb{j},\red{r}}+\hat{N}_{\vb{j},\green{g}}& 
    \hat{N}_{\vb{j},\text{pair}}&
    =\hat{N}_{\vb{j},\red{r}}\hat{N}_{\vb{j},\green{g}}\\
    \hat{N}_{\vb{j},\text{single}}&=\hat{N}_{\vb{j},\text{tot}}-\hat{N}_{\vb{j},\text{pair}}\\
\end{aligned}
\label{eq_number_operators2}
\end{equation}
which respectively measure the total matter density and the corresponding occupancy of pairs or single particles.

\paragraph{Dressed-site Casimir operator}
As we assumed via \cref{eq_Electric_casimir} that each of the two $\zeta$-rishons equally contributes to the link-electric energy density, we can equivalently define a dressed-site operator summing the Casimir contributions from its attached rishons:
\begin{equation}
    \hat{\Gamma}_{\vb{j}}=\frac{1}{2}\sum_{\vb*{\mu}}\hat{\vb{T}}_{\vb{j},\vb*{\mu}}^{2}
    \label{eq_gamma_operators}
\end{equation}

\paragraph{Corner Operators}
Expressing $\hat{U}$ as in \cref{eq_SU2_U_definition}, the plaquette interaction gives rise to 16 different terms:
\begin{equation*}
    \begin{split}
        U&_{\vb{j},\vb{j}+\vb*{\mu}_x}^{\alpha\beta}  
        {U_{\vb{j}+\vb*{\mu}_x,\vb{j}+\vb*{\mu}_x+\vb*{\mu}_y}^{\beta\gamma}}
        {U_{\vb{j}+\vb*{\mu}_x+\vb*{\mu}_y,\vb{j}+\vb*{\mu}_y}^{\gamma\delta}}
        {U_{\vb{j}+\vb*{\mu}_y,\vb{j}}^{\delta\alpha}}{=}\\
        &{=}
        \qty(\begin{array}{ccc}
        \ulcorner&\hat{\zeta}^{\delta\dagger}_{B,+\vb*{\mu}_{x}} \hat{\zeta}_{A,-\vb*{\mu}_{x}}^{\gamma}& \urcorner\\
        \hat{\zeta}_{A,-\vb*{\mu}_{y}}^{\delta}&  & \hat{\zeta}^{\gamma\dagger}_{B,-\vb*{\mu}_{y}}\\
        \hat{\zeta}^{\alpha\dagger}_{B,+\vb*{\mu}_{y}}&  &\hat{\zeta}_{A,+\vb*{\mu}_{y}}^{\beta}\\
        \llcorner& \hat{\zeta}_{A,+\vb*{\mu}_{x}}^{\alpha} \hat{\zeta}^{\beta\dagger}_{B,-\vb*{\mu}_{x}}&\lrcorner\\
        \end{array}) +\dots\\
        &\underset{*}{{=}}{-}\qty(\begin{array}{ccc}
        C^{AA}_{-\vb*{\mu}_{y},+\vb*{\mu}_{x}}&\rule{0.4cm}{0.5 pt}& C^{AA}_{-\vb*{\mu}_{x},-\vb*{\mu}_{y}}\\
        |&  &|\\
        C^{AA}_{+\vb*{\mu}_{x},+\vb*{\mu}_{y}}&\rule{0.4cm}{0.5 pt}&C^{AA}_{+\vb*{\mu}_{y},-\vb*{\mu}_{x}}\\
        \end{array}) +\dots,
    \end{split}
\end{equation*}
where, in $*$, we combined rishons in pairs to form \emph{corner operators} like the following:
\begin{equation}
\begin{aligned}
    \hat{C}^{AA}_{\vb{j},\vb*{\mu}_{1},\vb*{\mu}_{2}}&{=}\sum_{\alpha}
    \hat{\zeta}_{A,\vb{j},\vb*{\mu}_{1}}^{\alpha}\hat{\zeta}^{\alpha\dagger}_{A,\vb{j},\vb*{\mu}_{2}}\\
    &{=}\sum_{\alpha, \kappa, \kappa'}i\sigma^{y}_{\alpha,\kappa}
    \hat{\zeta}_{B,\vb{j},\vb*{\mu}_{1}}^{\kappa\dagger}i\sigma^{y}_{\alpha,\kappa'}\hat{\zeta}^{\kappa'}_{B,\vb{j},\vb*{\mu}_{2}}\\
    &{=}-\sum_{\alpha}
    \hat{\zeta}_{B,\vb{j},\vb*{\mu}_{1}}^{\alpha\dagger}\hat{\zeta}^{\alpha}_{B,\vb{j},\vb*{\mu}_{2}}=\hat{C}^{BB}_{\vb{j},\vb*{\mu}_{2},\vb*{\mu}_{1}}\\
    \hat{C}^{AB}_{\vb{j},\vb*{\mu}_{1},\vb*{\mu}_{2}}&{=}
    \sum_{\alpha}
    \hat{\zeta}_{A,\vb{j},\vb*{\mu}_{1}}^{\alpha}\hat{\zeta}^{\alpha\dagger}_{B,\vb{j},\vb*{\mu}_{2}}
    {=}\qty(\hat{C}^{BA}_{\vb{j},\vb*{\mu}_{2},\vb*{\mu}_{1}})^{\dagger}
\end{aligned}
\label{eq_corner_operators}
\end{equation}
As for the previous dressed-site operators, the practical construction of corner operators has to be consistent with the internal ordering in \cref{eq_SU2_dressed_site}.

\paragraph{SU(2) Link Symmetry}
Within the dressed-site formalism, the condition in \cref{SU2_linksymmetry} requiring the two rishon of the links to display the same Casimir operator is simply an Abelian Link Symmetry. It can be then easily encoded in TN libraries employing symmetries \cite{Silvi2019}.

\subsection{The operative defermionized Hamiltonian}
\begin{figure}
\centering
\includegraphics[width=1\columnwidth]{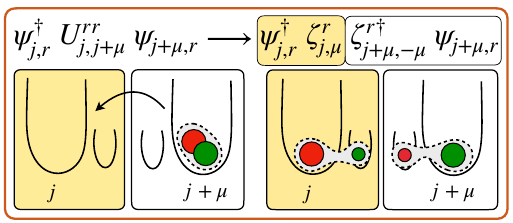}
\caption{Example of the matter-gauge interaction. 
Within the dressed-site formalism, the hopping always involves an even number of fermions (the matter field plus a rishon mode).}
\label{defermionization}
\end{figure}

We are then ready to rewrite the SU(2) lattice Yang-Mills Hamiltonian in \cref{eq_H_SU2_full} making use of dressed-site operators in Eq.s~\eqref{eq_arrival_operators}-\eqref{eq_corner_operators}. 
Namely, we have:
\begin{equation}
  \begin{split}
    &{\hat{H}}{=}{-}\frac{1}{2}\sum_{\vb{j}\in \Lambda}
    \Bigg[i\qty[\hat{Q}^{\dagger}_{A,\vb{j},+\vb*{\mu}_x}\hat{Q}_{B,\vb{j}+\vb*{\mu}_x,-\vb*{\mu}_x}{+}(A{\rightleftarrows} B)]\\
    &{+}({-}1)^{j_x + j_y}\qty[\hat{Q}^{\dagger}_{A,\vb{j},+\vb*{\mu}_y}\hat{Q}_{B,\vb{j}+\vb*{\mu}_y,-\vb*{\mu}_y}{+}(A{\rightleftarrows}B)]{+}{\text{H.c.}}\Bigg]\\
    &+ m\sum_{\vb{j}\in \Lambda} (-1)^{j_x + j_y} \hat{N}_{\vb{j},\text{tot}} + \frac{3 g^2}{16} \sum_{\vb{j}\in \Lambda} \hat{\Gamma}_{\vb{j}}\\
    &- \frac{8}{g^2}\sum_{\square \in \Lambda}\Re\qty(-\begin{matrix}
      \hat{C}^{AA}_{\ulcorner} &\hat{C}^{AA}_{\urcorner}\\
      \hat{C}^{AA}_{\llcorner} &\hat{C}^{AA}_{\lrcorner}
       \end{matrix}
       +\begin{matrix}
      \hat{C}^{BB}_{\ulcorner} &\hat{C}^{AB}_{\urcorner}\\
      \hat{C}^{AA}_{\llcorner} &\hat{C}^{AB}_{\lrcorner}
       \end{matrix}+\dots)
   \end{split}
   \label{H_full_pt2}
\end{equation}
Not surprisingly, \cref{H_full_pt2} is completely \emph{bosonic}, as all the dressed-site operators are made out of pairs of fermions (matter field + rishon, pairs of matter fields, or rishon pairs). 
Then, fermionic degrees of freedom are completely hidden inside each dressed site and there is no more need to face anti-commutation rules (see \cref{defermionization}).

Such an approach is inspired by \cite{Tagliacozzo2013, Silvi2014, Zohar2018, Zohar2019} and confirmed to be reliable when dealing with gauge theories and dynamical matter interacting in high-dimensional lattices. 
Therein, due to the presence of long-range strings of operators, the use of Jordan-Wigner transformation \cite{Jordan1928} or parity operators as in \cref{fermionic_qmb_op} is extremely inefficient from a numerical and experimental perspective. So far, only a few alternative techniques \cite{Verstraete2005, Corboz2010, Kraus2010} have been developed. 

\subsection{Projecting dressed site operators onto the gauge-invariant basis}
All the dressed-site operators obtained so far act on a Hilbert space of dimension 
\begin{equation}
    \text{dim} \mathcal{H}_{\text{dressed-site}}=\text{dim}\mathcal{H}_{\text{site}}\times \text{dim}\qty(\mathcal{H}_{\text{rishon}})^{2D},
\end{equation} 
where $D$ is the number of spatial dimensions.
Luckily, the subset of gauge invariant states is much smaller than $\mathcal{H}_{\text{dressed-site}}$. 
Therefore, the effective operators of the operative Hamiltonians should be obtained by projecting the obtained ones on the subspace generated by a gauge-invariant basis $M$.
Namely, for any dressed-site operator $\hat{O}$ among the previously defined, the corresponding effective operator $\hat{O}^{\text{eff}}$ acting on gauge-invariant states reads:
\begin{equation}
    \hat{O}^{\text{eff}}=M^{T}\cdot \hat{O}\cdot M
\end{equation}
where, for the gauge invariant basis, it holds:
\begin{align}
    M^{\dagger}\cdot M&=\mathbb{1} &
    \text{and}&&
    \qty(M\cdot M^{\dagger})^{2}&=\qty(M\cdot M^{\dagger}) 
\end{align}
As for the SU(2) LGT, the gauge-invariant basis $M$ is made by SU(2) singlets and can be determined as the kernel of the Gauss Law operators $\mathcal{G}_{SU(2)}$:
\begin{equation}
    \mathcal{G}_{\text{SU(2)}}\cdot M=\abs{\hat{\vb{S}}_{\text{matter}}{+}\sum_{\vb*{\mu}}\hat{\vb{T}}_{\vb*{\mu}}}^{2}\cdot M=0,
\end{equation}
where $\hat{\vb{T}}_{\vb*{\mu}}$ is the $\zeta$-rishon generator along the $\mu$ direction and defined in \cref{eq_T_generator}, while $\hat{\vb{S}}^{2}_{\text{matter}}$ is the matter color density introduced in \cref{eq_matter_Casimir}.
In the minimal $j_{\text{max}}=\frac{1}{2}$ truncation of SU(2), the gauge-invariant Hilbert space of every dressed site of the full Hamiltonian has 30 gauge invariant states, whereas, restricting to the pure theory, the local Hilbert space is 9-dimensional. 

More in general, the resulting operators $\hat{O}^{\text{eff}}$, for any spatial dimension D and any value of the gauge truncation $j_{\text{max}}$, are available in the GitHub repository \href{https://github.com/gcataldi96/ed-lgt}{ed-lgt}, which also allows for simulations via Exact Diagonalization. 

\section{Non-local/topological properties}
\label{app_topology}
In this section, we address the topological properties of the (2+1)D minimally truncated SU(2) Yang-Mills LGT. 
In particular, we show that the pure theory in \cref{eq_H_SU2_pure} displays a non-local $\mathbb{Z}_{2}\times\mathbb{Z}_{2}$ symmetry whose topological sectors closes as approaching the $g$-transition. 
Such a topological structure disappears in the full Hamiltonian \cref{eq_H_SU2_full} but can be recovered in the infinite mass limit.
We stress that the topological symmetry is completely \emph{independent} of the chosen truncation of the SU(2) gauge Hilbert space developed throughout \cref{app_model}.

Let us start searching for some topological invariants. 
The right candidates involve the rishon parity operators $\hat{P}_{\zeta}$ introduced in \cref{eq_SU2_rishon_parity}. 
Thanks to the link symmetry in \cref{SU2_linksymmetry}, we can extend such a definition to the whole link $(\vb{j},\vb{j}+\vb*{\mu})$ and consider the corresponding parity operator $\hat{P}_{\vb{j},\vb{j}+\vb*{\mu}}$. 
As aforementioned, it returns ($+1$) for \emph{integer} and ($-1$) for \emph{semi-integer} SU(2) representations. 
In our 5-dimensional $(0{\otimes} 0){\oplus}(\frac{1}{2}{\otimes}\frac{1}{2})$ SU(2) minimally truncated Hilbert space, such an operator reads:
\begin{equation}
    \hat{P}_{\vb{j},\vb{j}+\vb*{\mu}} =\qty(\begin{array}{c|cccc}
      +1&0&0&0&0\\
      \hline
      0&-1&0&0&0\\
      0&0&-1&0&0\\
      0&0&0&-1&0\\
      0&0&0&0&-1\\
  \end{array})\quad \forall \vb{j}, \, \forall \vb*{\mu}
  \label{link_Parity}
\end{equation}
By definition, $\hat{P}_{\vb{j},\vb{j}+\vb*{\mu}}$ commutes with the Casimir operator in \cref{Casimir2}, as both are diagonal in the link basis:
\begin{equation}
    \qty[\hat{P}_{\vb{j},\vb{j}+\vb*{\mu}}, \hat{E}^{2}_{\vb{j},\vb{j}+\vb*{\mu}}]=0\quad \forall \vb{j},\, \forall \vb*{\mu}
\end{equation}
Rather, $\hat{P}_{\vb{j},\vb{j}+\vb*{\mu}}$ anti-commutes with the parallel transport $\hat{U}$ (and $\hat{U}^{\dagger}$), as its action on the link decreases (respectively increases) the SU(2) link-representation by $1/2$:
\begin{align}
    \qty{\hat{P}_{\vb{j},\vb{j}+\vb*{\mu}}, \hat{U}^{(\dagger)}_{\vb{j},\vb{j}+\vb*{\mu}}}&=0&\forall \vb{j},\, \forall \vb*{\mu}&\in \Lambda
\end{align}

\begin{figure}
	\centering
	\includegraphics[width=1\columnwidth]{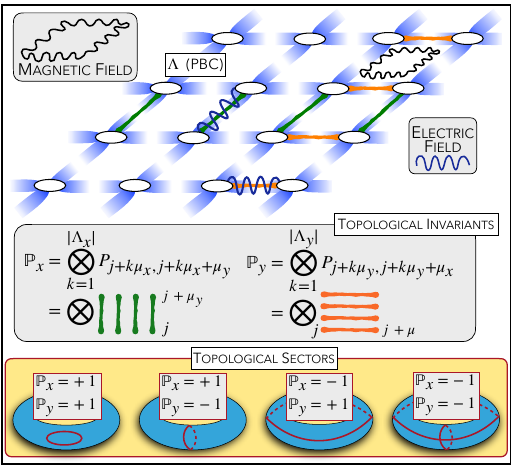}
	\caption{Pictorial representations of the topological invariants defined in \cref{Py} and \cref{Px} on a lattice $\Lambda$ in PBC (i.e. a torus). The topological sectors of \cref{topological_sectors} are sketched in the yellow panel: closed red curves on the blue torus $\Lambda$ correspond to SU(2) loop excitations.}
	\label{topology}
\end{figure}

Then, let us consider our 2D lattice $\Lambda$ in PBC and introduce the consecutive action of the horizontal link parity operators along a vertical loop in $\Lambda$ (see orange links in \cref{topology}). Namely, we define:
\begin{equation}
    \begin{split}
\mathbb{\hat{P}}_{y}&\equiv\bigotimes_{k=0}^{\abs{\Lambda_{y}}}\hat{P}_{\vb{j}+k\vb*{\mu}_{y},\vb{j}+k\vb*{\mu}_{y}+\vb*{\mu}_{x}}\\
&=\hat{P}_{\vb{j},\vb{j}+\vb*{\mu}_{x}}\otimes \hat{P}_{\vb{j}+\vb*{\mu}_{y},\vb{j}+\vb*{\mu}_{y}+\vb*{\mu}_{x}}\otimes \dots
    \end{split}
    \label{Py}
\end{equation}
Correspondingly, the consecutive action of the vertical link parity operator along a horizontal loop in $\Lambda$ (see green links in \cref{topology}) is
\begin{equation}
    \begin{split}
\hat{\mathbb{P}}_{x}&\equiv\bigotimes_{k=0}^{\abs{\Lambda_{x}}}\hat{P}_{\vb{j}+k\vb*{\mu}_{x},\vb{j}+k\vb*{\mu}_{x}+\vb*{\mu}_{y}}\\
&=\hat{P}_{\vb{j},\vb{j}+\vb*{\mu}_{y}}\otimes \hat{P}_{\vb{j}+\vb*{\mu}_{x},\vb{j}+\vb*{\mu}_{x}+\vb*{\mu}_{y}}\otimes \dots
    \end{split}
    \label{Px}
\end{equation}

It is clear that both the $\hat{\mathbb{P}}_{x}$ and $\hat{\mathbb{P}}_{y}$ operators remain unaffected by the action of the electric field along any of their steps, as their parity does not get flipped. 
Correspondingly, any plaquette term $\hat{B}^{2}_{\square}$ of the magnetic interaction applied on the chain where $\hat{\mathbb{P}}_{x}$ or $\hat{\mathbb{P}}_{y}$ is evaluated flips the parity of two consecutive steps of the chain so that the overall sign is left unchanged.
 Namely:
\begin{align}
\qty[\hat{\mathbb{P}}_{\vb{j}},\hat{E}^{2}_{\vb{j},\vb{j}+\vb*{\mu}}]=0
=\qty[\hat{\mathbb{P}}_{\vb{j}},\hat{B}^{2}_{\square}] \qquad \forall \vb{j}, \square \in \Lambda
\end{align}
We conclude that $\hat{\mathbb{P}}_{x}$ and $\hat{\mathbb{P}}_{y}$ are generators of two symmetries of the pure Hamiltonian in \cref{eq_H_SU2_pure}:
\begin{align}
    \qty[\hat{\mathbb{P}}_{x},\hat{H}_{\text{pure}}]&=0 &
    \qty[\hat{\mathbb{P}}_{y},\hat{H}_{\text{pure}}]&=0
\end{align}
and we can refer to them as \emph{topological invariants}.
The whole symmetry group is then $\mathbb{Z}_{2}\times\mathbb{Z}_{2}$, as we have $\qty[\hat{\mathbb{P}}_{x},\hat{\mathbb{P}}_{y}]=0$ $\forall x,y\in \Lambda$. 
Therefore, any physical state $\ket{\Psi}$ of the pure theory in \cref{eq_H_SU2_pure} lies in one of the sectors of $\hat{\mathbb{P}}_{x}$ and $\hat{\mathbb{P}}_{y}$ sketched in the yellow panel of \cref{topology}. 
The distinction between different symmetry sectors is in terms of the number of \emph{non-removable loop excitations} displayed by the state. 
With \emph{loop excitations}, we refer to closed magnetic strings (red circles in the blue torus of the yellow panel of \cref{topology}) displayed by the state on its topological geometry. 
In particular, \emph{non-removable} loops are the ones that cannot be removed through homotopies, i.e. without modifying the topology of the system.

Then, any state with an \emph{even} number of horizontal (vertical) non-removable loop excitations lies in the \emph{even} sector of the vertical $\hat{\mathbb{P}}_{y}$ (horizontal $\hat{\mathbb{P}}_{x}$) topological invariant. Correspondingly, any state with an \emph{odd} number of non-removable loop excitations lies in the \emph{odd} sector of the proper topological invariant.
Hence, $\forall k \in \qty{x,y}$:
\begin{align}
    \expval{\hat{\mathbb{P}}_{k}}{\Psi}&{=}\lambda &\text{where}&& \lambda& \in
    \begin{array}{c|cccc}
        \hat{\mathbb{P}}_{x}&+1 &+1&-1&-1 \\
        \hline
        \hat{\mathbb{P}}_{y}&+1 &-1&+1&-1 \\
    \end{array}
    \label{topological_sectors}
\end{align}
Such symmetry explicitly disappears in the full Hamiltonian \cref{eq_H_SU2_full} because of the hopping terms, as each of them includes a single parallel transport $\hat{U}$ that flips one link parity along the line where $\mathbb{P}_{x}$ or $\hat{\mathbb{P}}_{y}$ are defined. However, in the large-$m$ limit, where the full Hamiltonian \cref{eq_H_SU2_full} falls back into the pure theory in \cref{eq_H_SU2_pure}, we expect to recover the same topological invariants (at least in the ground-state).

To check numerically the previous statements, we would need to measure the topological invariants on the low energy states of the Hamiltonian in \cref{H_full_pt2}. 
Within our dressed site formalism, \cref{Py} and \cref{Px} can be expressed just as chains of single-site operators along one of the two sides of the links. 
Indeed, as long as the SU(2) link-symmetry in \cref{SU2_linksymmetry} is satisfied, the information about every link-parity is present in both the attached neighboring sites. 

As shown in \cref{fig_topological_sectors}, the topological sectors of the first 4 lowest eigenstates of the pure theory in PBC belong to a different topological sector of \cref{topological_sectors}.
Moreover, the eigenstates are sorted in increasing energy according to the table in \cref{topological_sectors}. 
In particular, $E_{1}=E_{2}$ only in the case of isotropic geometries, as non-removable loop excitations along the two directions are equally expensive in energy. 
In the case of an-isotropic lattices, where $\abs{\Lambda}_{x}\neq \abs{\Lambda}_{y}$, non-removable loop excitations along different axes are shifted in energy.

As for the full theory, we restrict our simulations to the zero charge density sector of a $2\times2$ lattice in PBC with $m\in\qty[10^{-2},10^{+2}]$. 
\begin{figure}
	\centering
	\includegraphics[width=1\columnwidth]{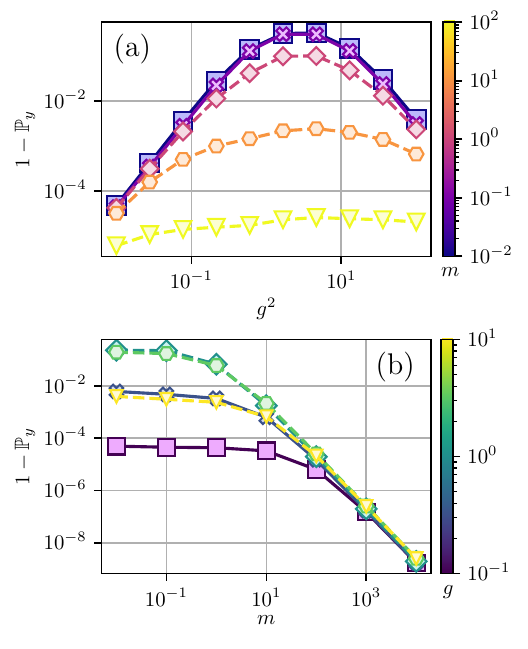}
	\caption{Distance between the ground-state $\mathbb{P}_{y}$-topological invariant in the full theory and the corresponding one of the pure theory for different $m$-values (a) and $g$-couplings (b). Results from simulations in a $2\times 2$ lattice with PBC at $b=0$.}
	\label{fig_full_topology}
\end{figure}

In \cref{fig_full_topology}, we look at the distance between the exact \emph{even} topological sector of $\hat{\mathbb{P}}_{y}$ (+1) and the corresponding value measured on the ground state. 
That gap gets larger when approaching the $g$-transition while vanishing far from the latter. Moreover, as aforementioned, in the large-mass $m$ limit, we recover the full symmetry sector of the pure theory.

\section{Large-$g$ phase via perturbation theory}
\label{app_large_g_perturbation_theory}
In the large-$g$ limit, the zero-density sector of the truncated SU(2) Hamiltonian can be studied via perturbation theory. 
In this regime, we can rewrite \cref{H_full_pt2} as 
\begin{equation}
\begin{split}
    H&{\sim}\qty[\hat{H}_{0}{+}\qty(\hat{H}_{\text{matter}}{+}\hat{H}_{\text{x-hop}}{+}\hat{H}_{\text{y-hop}}{+}\hat{H}_{\text{plaq}})]\\
    &{=}\sum_{\vb{j}\in \Lambda}\qty[\hat{h}^{0}_{\vb{j}}{+}\hat{h}^{\text{matter}}_{\vb{j}}{+}\hat{h}^{\text{x-hop}}_{\vb{j}}{+}\hat{h}^{\text{y-hop}}_{\vb{j}}{+}\hat{h}^{\text{plaq}}_{\vb{j},\square}]
\end{split}
\end{equation}
where $\hat{h}_{0}$ is the single-site unperturbed Hamiltonian:
\begin{equation}
  \begin{aligned}
    \hat{h}^{0}_{\vb{j}}=\frac{3g^{2}}{16}\hat{\Gamma}_{\vb{j}} 
  \end{aligned} 
  \label{H0_perturbative}
\end{equation}
while the perturbative terms read:
\begin{align}
\hat{h}^{\text{matter}}_{\vb{j}}&=m(-1)^{j_x + j_y} \hat{N}_{\vb{j},\text{tot}}\label{H_matter}\\
\hat{h}^{\text{x-hop}}_{\vb{j}}&=\frac{1}{2}
    \qty[-i \hat{Q}^{\dagger}_{\vb{j},+\vb*{\mu}_x}\hat{Q}_{\vb{j}+\vb*{\mu}_x,-\vb*{\mu}_x}+ \text{H.c.}]\label{xhop}\\
\hat{h}_{\vb{j}}^{\text{y-hop}}&=\frac{1}{2}\qty[- (-1)^{j_x+j_y}\hat{Q}^{\dagger}_{\vb{j},+\vb*{\mu}_y} \hat{Q}_{\vb{j}+\vb*{\mu}_y,-\vb*{\mu}_y}+ \text{H.c.}]\label{yhop}\\
\hat{h}^{\text{plaq}}_{\vb{j},\square}&=
    - \frac{8}{g^{2}}\qty(\begin{matrix}
    \hat{C}_{\ulcorner} &\hat{C}_{\urcorner}\\
    \hat{C}_{\llcorner} &\hat{C}_{\lrcorner}
    \end{matrix}
    )
    \label{Hplaq}
\end{align}
where for simplicity, we replaced the two species of \emph{arrival operators} in \cref{eq_arrival_operators} and \emph{corner operator} in \cref{eq_corner_operators} with a unique version, $\hat{Q}$ and $\hat{C}$ respectively.

In the large-$g$ limit, we expect the $0^{th}$ order ground-state $\ket{E_{0}}$ not to display gauge activity, as the electric interaction is energetically penalized. 
Then, the effective Hilbert state of the dressed sites reduces just to states with singlets in the matter fields. 
Namely, in terms of the sectors of the local charge density operator, we have only
\begin{align}
   \ket{0}&\equiv \five{0}{0}{0}{0}{0} &
   \text{and}&&
   \ket{2}&\equiv \five{\red{r}\green{g}}{0}{0}{0}{0}
   \label{02_singlesite_states}
\end{align}
Therefore, at the $0^{th}$-order, the single-site ground-state $\ket{E_{0}}$ can be expressed as a linear combination of \cref {02_singlesite_states} with energy $E_{0}=0$:
\begin{equation}
\ket{E_{0}}=\alpha\ket{0}+\beta\ket{2} \qquad \text{with}\qquad \sqrt{\alpha^{2}+\beta^{2}}=1
\label{ground_state_order0}
\end{equation}
\begin{figure}
	\centering
	\includegraphics[width=1\columnwidth]{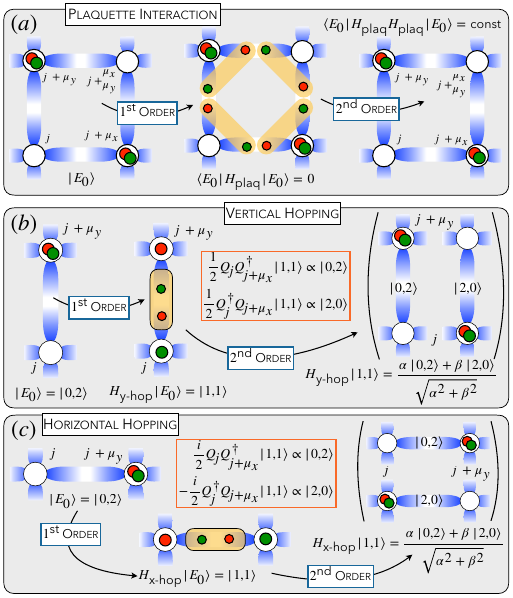}
	\caption{Graphical representation of the $1^{st}$ and $2^{nd}$ order perturbative effects of the magnetic and the hopping terms to the ground state of \cref{H0_perturbative}.}
	\label{perturbative_figure}
\end{figure}
At the $1^{st}$ perturbative order, we have to separately consider the action of every single term in \cref{H_matter}-\eqref{Hplaq}. 
As for the plaquette term in \cref{Hplaq}, we expect it to yield a vanishing contribution. Indeed, if we refer to $\ket{E_{0}}$ as a single-plaquette ground state, then we have:
\begin{equation}
\begin{split}
    \bra{E_{0}}\hat{h}^{\text{plaq}}\ket{E_{0}}
    &=\bra{E_{0}}\qty(-\frac{8}{g^{2}})\qty(\frac{1}{\sqrt{2}})^{4}\ket{E_{1}}\\&=-\frac{2}{g^{2}}\braket{E_{0}}{E_{1}}=0
\end{split}
\end{equation}
since the plaquette-state $\ket{E_{1}}$ is orthogonal to the ground state $\ket{E_{0}}$, as all its links are electrically active (see \cref{perturbative_figure}).
The factor $1/\sqrt{2}$ is due to each single corner operator $\hat{C}$ acting on the corresponding empty corner of the plaquette state $\ket{E_{0}}$.  

As for the hopping terms, we focus on the effective Hilbert space of the joint neighboring sites $\vb{j}$ and $\vb{j}+\vb*{\mu}$:
\begin{equation}
\mathcal{H}^{\text{eff}}_{\vb{j},\vb{j}+\vb*{\mu}}=
\qty{\ket{0,0},\ket{0,2},\ket{2,0},\ket{2,2}}\qquad \forall \vb{j},\,\forall \vb*{\mu}
\end{equation}
where we labeled the states $\ket{\vb{j},\vb{j}+\vb*{\mu}}$ in terms of the only two possible single-site states in \cref{02_singlesite_states}. First of all, we notice that $\ket{0,0}$ and $\ket{2,2}$ are completely decoupled from the other two states, since $\forall \vb*{\mu}$:
\begin{equation}
\hat{Q}^{\dagger}_{\vb{j},+\vb*{\mu}}
\hat{Q}_{\vb{j}+\vb*{\mu},-\vb*{\mu}}\ket{0,0}=
\hat{Q}^{\dagger}_{\vb{j},+\vb*{\mu}}
\hat{Q}_{\vb{j}+\vb*{\mu},-\vb*{\mu}}\ket{2,2}=0
\label{hop_decoupled_states}
\end{equation}
Then, the only relevant matrix elements of the effective (perturbed) hopping-Hamiltonian are the following ones:
\begin{equation}
    \begin{aligned}
\hat{Q}^{\dagger}_{\vb{j},+\vb*{\mu}}
\hat{Q}_{\vb{j}+\vb*{\mu},-\vb*{\mu}}\ket{0,2}=&(-1)^{2}\ket{1,1}\\
\hat{Q}_{\vb{j},+\vb*{\mu}}
\hat{Q}^{\dagger}_{\vb{j}+\vb*{\mu},-\vb*{\mu}}\ket{2,0}=&(-1)^{2}\ket{1,1}
    \end{aligned}
    \qquad \forall \vb{j},\,\forall \vb*{\mu}
    \label{hop_coupled_states}
\end{equation}
where $\ket{1,1}$ is figured in \cref{perturbative_figure}, while the $(-1)$ factor is due to the action of a single arrival operator $\hat{Q}_{\vb{j},\vb*{\mu}}^{(\dagger)}$ defined in \cref{eq_arrival_operators} on the states in \cref{02_singlesite_states}. 
Clearly, since 
\begin{equation}
    \braket{0,2}{1,1}=0= \braket{2,0}{1,1},
\end{equation}
none of the hopping Hamiltonians \cref{xhop}-\eqref{yhop} do provide any $1^{st}$-order correction to $\hat{H}_{0}$ in \cref{H0_perturbative}.

The only relevant $1^{st}$ order term is the one related to $\hat{H}_{\text{matter}}$, as it acts just on the matter fields without yielding any gauge activity. 
Moreover, it removes the ground-state degeneracy of \cref{ground_state_order0} by favoring a staggered configuration to the lattice, namely:
\begin{equation}
\ket{E_{1}(\vb{j})}=\delta_{1,(-1)^{\vb{j}_{x}+\vb{j}_{y}}}\ket{0}+\delta_{-1,(-1)^{\vb{j}_{x}+\vb{j}_{y}}}\ket{2}
\label{ground_state_order1}
\end{equation}
where $\delta_{ij}$ is the Kronecker delta function.
However, for sufficiently small values of the mass $m$, the staggering effect is irrelevant, and the degeneracy of \cref{ground_state_order0} is restored. 
Therefore, in the small-$m$ limit, the first relevant perturbative order is the $2^{nd}$ one. 

As for the plaquette interaction, the $2^{nd}$ order does not remove the ground-state degeneracy, as it completely restores $\ket{E_{0}}$ providing just an energy shift.
Namely, the $2^{nd}$ order perturbative corrections to the single-site ground-state energy reads:
\begin{equation}
\begin{split}
E_{2}^{\text{plaq}}&=\frac{1}{4}\expval{\hat{h}_{\text{plaq}}
[E_{0}-H_{0}]^{-1}\hat{h}_{\text{plaq}}}{E_{0}}\\
&=\frac{1}{4}\qty(-\frac{2}{g^{2}})\bra{E_{0}} \hat{h}_{\text{plaq}}[E_{0}-H_{0}]^{-1}\ket{E_{1}}\\
&=-\frac{1}{2g^{2}}\bra{E_{0}} \hat{h}_{\text{plaq}}\qty[-\frac{3g^{2}}{16}\sum_{\vb{j}\in\square}\hat{\Gamma}_{\vb{j}}]^{-1}\ket{E_{1}}\\
&=-\frac{1}{2g^{2}}\qty(-8\cdot\frac{3g^{2}}{16})^{-1}\bra{E_{0}} \hat{h}_{\text{plaq}}\ket{E_{1}}\\
&=-\frac{1}{3g^{4}}\qty(-\frac{2}{g^{2}})\braket{E_{0}}=\boxed{\frac{2}{3g^{6}}}
\end{split}
\end{equation}
where $[\hat{O}]^{-1}$ is the Moore-Penrose inverse and the initial $1/4$ factor is put to get the single-site energy out of the one of a plaquette.

As for the hopping terms, because of \cref{hop_decoupled_states}-\eqref{hop_coupled_states}, $\forall k\in\qty{x,y}$, the only relevant terms are the diagonal ones
\begin{equation}
\begin{split}
    \frac{1}{2}&\bra{0,2}\hat{h}^{\text{hop}}[E_{0}-\hat{H}_{0}]^{-1}\hat{h}^{\text{hop}}\ket{0,2}\\
    &=\frac{1}{2}\bra{2,0}\hat{h}^{\text{hop}}[E_{0}-\hat{H}_{0}]^{-1}\hat{h}^{\text{hop}}\ket{2,0}
\end{split}
\end{equation}
and the off-diagonal ones:
\begin{equation}
\begin{split}
    \frac{1}{2}&\bra{0,2}\hat{h}^{\text{hop}}[E_{0}-\hat{H}_{0}]^{-1}\hat{h}^{\text{hop}}\ket{2,0}\\
    &=\frac{1}{2}\bra{2,0}\hat{h}^{\text{hop}}[E_{0}-\hat{H}_{0}]^{-1}\hat{h}^{\text{hop}}\ket{0,2}\\
\end{split}
\end{equation}
The factor $1/2$ is put to take into account just the single-site energy out of the corresponding two-site energy. As for the hopping along the $x$-axis, we have:
\begin{equation}
    \begin{split}
    &{\frac{1}{2}}\bra{0,2}\hat{h}^{\text{x-hop}}
    [E_{0}-\hat{H}_{0}]^{-1}\hat{h}^{\text{x-hop}}\ket{0,2}\\
    &{=}\frac{1}{2}\bra{0,2}h^{\text{x-hop}}[E_{0}-\hat{H}_{0}]^{-1}\qty(-\frac{i}{2})\ket{1,1}\\
    &{=}\qty(\frac{-i}{4})\bra{0,2}\hat{h}^{\text{x-hop}}\qty[-\frac{3g^{2}}{16}\qty(\hat{\Gamma}_{\vb{j}}{+}\hat{\Gamma}_{\vb{j}+\vb*{\mu}_{x}})]^{-1}\ket{1,1}\\
    &{=}\qty(\frac{-i}{4})\bra{0,2}\hat{h}^{\text{x-hop}}\qty(-\frac{8}{3g^{2}})\ket{1,1}\\
    &{=}\frac{2i}{3g^{2}}\bra{0,2}\qty(\frac{i}{2})\ket{0,2}=\boxed{-\frac{1}{3g^{2}}}
    \end{split}
\end{equation}
Analogously proceeding, we have:
\begin{equation}
    \begin{split}
    &\frac{1}{2}\bra{2,0}\hat{h}^{\text{x-hop}}[E_{0}-\hat{H}_{0}]^{-1}\hat{h}^{\text{x-hop}}\ket{0,2}=\boxed{\frac{1}{3g^{2}}}
    \end{split}
\end{equation}
Then, the $2^{nd}$-order perturbative $x$-hopping term reads:
\begin{equation}
\begin{split}
    \hat{h}_{\text{x-hop}}^{\text{eff}}&=-\frac{1}{3g^{2}}
    \begin{pmatrix}
    0&0&0&0\\
    0&+1&-1&0\\
    0&-1&+1&0\\
    0&0&0&0\\
    \end{pmatrix}\\
    &=-\frac{1}{6g^{2}}\qty[\hat{\sigma}^{x}_{\vb{j}}\hat{\sigma}^{x}_{\vb{j}+\vb*{\mu}_{y}}+\hat{\sigma}^{y}_{\vb{j}}\hat{\sigma}^{y}_{\vb{j}+\vb*{\mu}_{y}}-\hat{\sigma}^{z}_{\vb{j}}\hat{\sigma}^{z}_{\vb{j}+\vb*{\mu}_{y}}]
\end{split}
\end{equation}
As for the $y$-hopping Hamiltonian, one can prove that:
\begin{equation}
\begin{split}
    \hat{h}_{\text{y-hop}}^{\text{eff}}&=\frac{1}{3g^{2}}
    \begin{pmatrix}
    0&0&0&0\\
    0&+1&+1&0\\
    0&+1&+1&0\\
    0&0&0&0\\
    \end{pmatrix}\\
    &=-\frac{1}{6g^{2}}\qty[\hat{\sigma}^{x}_{\vb{j}}\hat{\sigma}^{x}_{\vb{j}+\vb*{\mu}_{y}}+\hat{\sigma}^{y}_{\vb{j}}\hat{\sigma}^{y}_{\vb{j}+\vb*{\mu}_{y}}+\hat{\sigma}^{z}_{\vb{j}}\hat{\sigma}^{z}_{\vb{j}+\vb*{\mu}_{y}}]
\end{split}
\end{equation}
Summarizing, in the large-$g$ limit, the Hamiltonian in \cref{H_full_pt2} can be approximated as:
\begin{equation}
  \begin{aligned}
    \hat{H}^{\text{eff}}{\sim}&{-}{\frac{1}{6g^{2}}}\sum_{\vb{j},\vb*{\mu}_{x}}
    \qty[\hat{\sigma}^{x}_{\vb{j}}\hat{\sigma}^{x}_{\vb{j}+\vb*{\mu}_{x}}{+}\hat{\sigma}^{y}_{\vb{j}}\hat{\sigma}^{y}_{\vb{j}+\vb*{\mu}_{x}}{-}\hat{\sigma}^{z}_{\vb{j}}\hat{\sigma}^{z}_{\vb{j}+\vb*{\mu}_{x}}]\\
    &{-}{\frac{1}{6g^{2}}}\sum_{\vb{j},\vb*{\mu}_{y}}
    \qty[\hat{\sigma}^{x}_{\vb{j}}\hat{\sigma}^{x}_{\vb{j}+\vb*{\mu}_{y}}{+}\hat{\sigma}^{y}_{\vb{j}}\hat{\sigma}^{y}_{\vb{j}+\vb*{\mu}_{y}}{+}\hat{\sigma}^{z}_{\vb{j}}\hat{\sigma}^{z}_{\vb{j}+\vb*{\mu}_{y}}]\\
    \end{aligned}
\end{equation}
which looks similar to a 2D quantum Heisenberg Hamiltonian apart from the staggering factor in the kinetic term $\hat{\sigma}^{z}\hat{\sigma}^{z}$.

\section{Exact results at small sizes}
\label{app_ED_results}
As pointed out in \cref{results}, addressing the 2D SU(2) LGT on large system sizes is significantly demanding, especially in the small-$g$ (magnetic) phase and close to the $g$-transition, because of the large entanglement displayed by the model. 
Nevertheless, by exploiting small system simulations at maximum TTN bond dimension (equivalent to ED), we can provide compelling features of both the pure and the full SU(2) theories. 

\begin{figure}
	\centering
	\includegraphics[width=1\columnwidth]{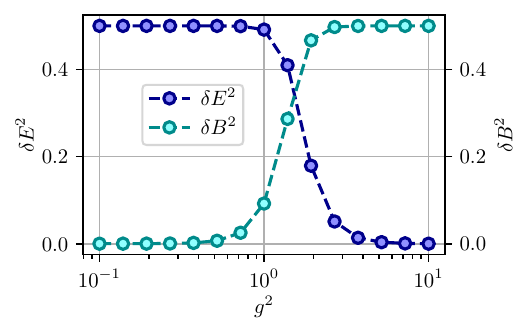}	\caption{Quantum fluctuations of the gauge observables in \cref{gauge_observables1}-\eqref{gauge_observables2} as a function of the $g$-coupling. Results obtained from simulations of a $2\times 2$ lattice in OBC.}
	\label{pure_fluctuations}
\end{figure}

As for the pure theory, we show in \cref{pure_fluctuations} that the magnetic and the electric phases discussed in \cref{sec_pure_theory} are characterized by strong fluctuations
\begin{align}
  \delta E^{2}&=\sqrt{\avg{E^4} - \avg{E^2}^2}&
  \delta B^{2}&=\sqrt{\avg{B^4} - \avg{B^2}^2}
\end{align}
of the gauge observables in \cref{gauge_observables1}-\eqref{gauge_observables2} respectively. 
In particular, the magnetic (small-$g$) phase is characterized by large and strong fluctuating electric energy, while the electric (large-$g$) phase displays large and strong fluctuating magnetic energy.

As for the full SU(2) theory, to select the chosen baryon density sector $b^{*}$ 
we add to \cref{H_full_pt2} the term
\begin{equation}
\hat{H}_{b} = \Tilde{\eta} \qty(\sum_{\vb{j}\in \Lambda}\sum_{\alpha}\hat{\psi}^{\dagger}_{\vb{j},\alpha}\hat{\psi}_{\vb{j},\alpha} +2 -2b^{*})^2
\end{equation}
where $\Tilde{\eta}$ plays the role of a large penalty coefficient that increases the energy associated with baryon number densities differing from $b^{*}$. When exploiting TTN methods, the chosen symmetry sector is externally selected by directly encoding the abelian symmetry $U(1)$ in the TTN ansatz \cite{Singh2010, Singh2011, Orus2019, Silvi2019}.

\begin{figure*}
	\centering
	\includegraphics[width=1\textwidth]{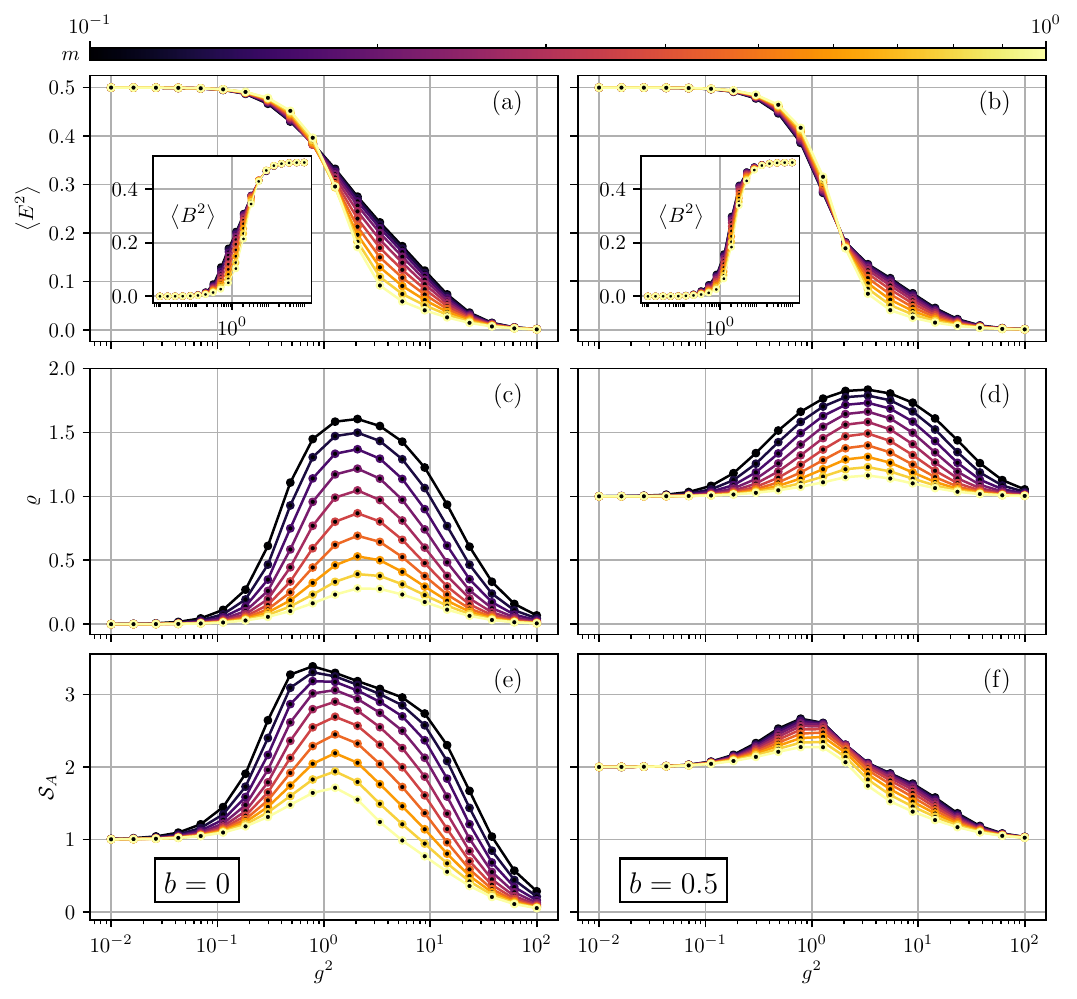}
	\caption{Simulations of the full SU(2) Hamiltonian in \cref{H_full_pt2} on a $2\times 2$ lattice with OBC in the $b=0$ (a)-(c)-(e) and $b=0.5$ (b)-(d)-(f) baryon number density sectors. 
    The plots display respectively: (a)-(b) the average electric and magnetic energy contributions $\avg{E^{2}}$ and $\avg{B^{2}}$ (inset) enlightening the transition between the magnetic and the electric phases discussed in \cref{sec_pure_theory}; (c)-(d) the average particle density $\varrho$ in \cref{density}, which appears peaked in the $g$-transition; (e)-(f) the entanglement entropy $\mathcal{S}_{A}$ of half the system, with a peak in the $g$-transition which is larger for smaller $m$ while disappearing for large ones.
        }
	\label{fig_ED_SU2_grid_discrete}
\end{figure*}

In \cref{fig_ED_SU2_grid_discrete}, we focus on the local observables in \cref{gauge_observables1}-\eqref{gauge_observables2}-\eqref{density} and the entanglement entropy in \cref{entropy} 
under OBC at $b=0$ and $b=0.5$. 
By varying $m\in\qty[10^{-1},10^{0}]$, we notice that, for both the baryon-density sectors, the larger the mass $m$, the sharper the transition between the \emph{magnetic} (small-$g$) and the \emph{electric} (large-$g$) phases. 

\begin{figure*}
	\centering
\includegraphics
[width=1\textwidth]{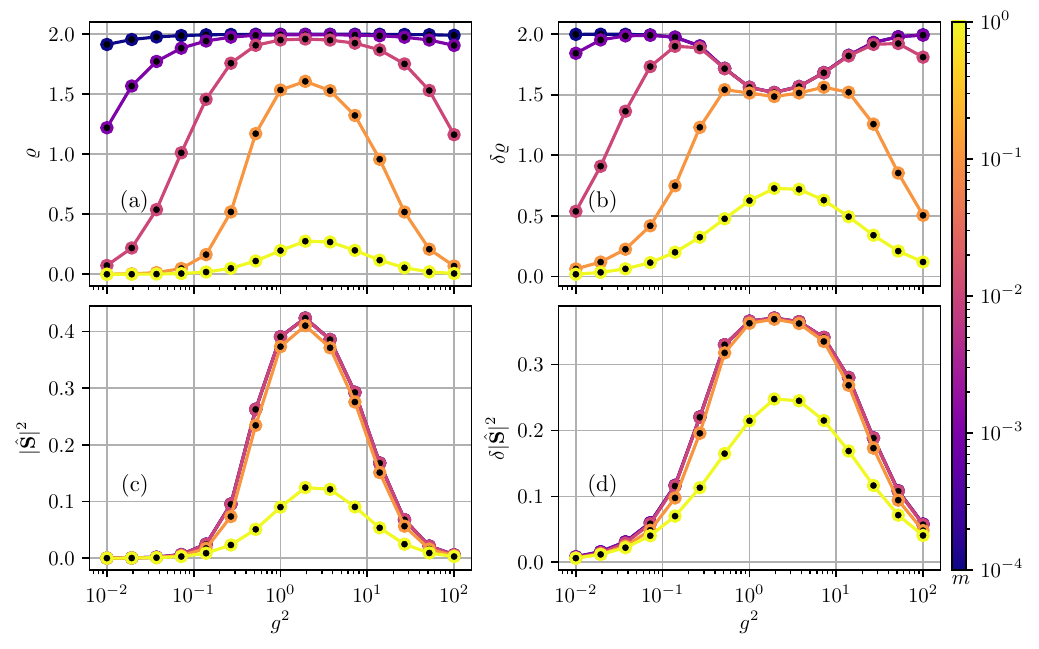}
	\caption{Simulations of a $2\times2$ lattice in PBC. The plots display respectively: (a) the average particle density $\rho$ and (b) its quantum fluctuations $\delta \rho$; (c) the matter color density $\abs{\vb{S}}^{2}$ and (d) its quantum fluctuations $\delta \abs{\vb{S}}^{2}$. All the observables are studied as a function of the square coupling $g^{2}$ for different mass values $m\in \qty[10^{-4},10^{1}]$.}
\label{fig_low_mass_behavior}
\end{figure*}

Moreover, as discussed in \cref{sec_full_theory}, in between the two phases, the model is characterized by a \emph{baryon-liquid phase}, where the particle density $\varrho$ defined in \cref{density} reveals peaked and strong fluctuating (check also \cref{fig_low_mass_behavior}(a)-(b)), and the peak is higher and larger for smaller $m$-values.

Correspondingly, for fixed $m$-values, the entanglement entropy of half the lattice is constant in the magnetic phase, peaked in the $g$-transition, and tending to a default value ($0$ for $b=0$ and $1$ for $b=0.5$) in the electric phase. 
In the limit of large masses, this peak in the entropy progressively vanishes, and we recover the (\emph{crossover}/\emph{first-order}) transition observed in \cref{pure_theory_simulations} for the pure theory. 

Within the Baryon-liquid phase, we can also notice a non-null value of the color density defined in \cref{eq_matter_Casimir}, which is related to the presence of single particle non-null expectation values. 
As shown in \cref{fig_low_mass_behavior}, $|\hat{\vb{S}}|^{2}$ displays a peak in the proximity of the magnetic-electric $g$-transition and is supported by fluctuations $\delta|\hat{\vb{S}}|^{2}$ of the same order of magnitude of the observable itself. 
In these terms, this phase represents the only candidate for displaying deconfinement. At any rate, it does not survive in the continuum limit, as it remains bound in intermediate $g$-values. 

\section{Tensor Network Methods}
\label{app_tensor_networks}
Tensor network simulations performed in this work have used the Tree Tensor Networks ground state variational searching algorithm \cite{Silvi2019, Cataldi2021, Ferrari2022}.
Given the Hamiltonian of the considered system, the unconstrained binary TTN is constructed, and the ground state is determined by optimizing all the tensors in the tree network with a fixed bond dimension $\chi$. 

\begin{figure*}
	\centering
	\includegraphics[width=1\textwidth]{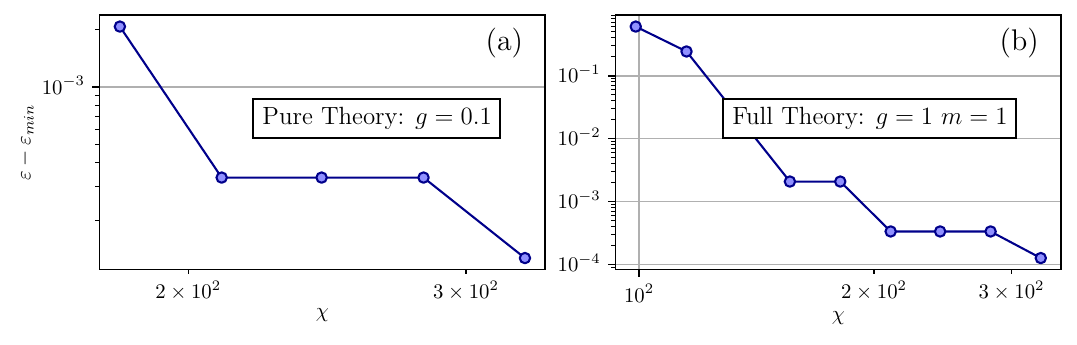}
	\caption{Convergence tests for (a) the pure and the (b) full Hamiltonian in \cref{H_full_pt2} on a $4\times4$ lattice in OBC. 
 The plots display the scaling of the ground-state energy density $\varepsilon$ (up to its minimal value $\varepsilon_{min}$) as a function of the bond dimension $\chi$ adopted in the TTN simulations.}
	\label{fig_conv_tests}
\end{figure*}

In detail, we exploit the Krylov sub-space expansion technique to numerically solve the local eigenvalue problem for each tensor \cite{Silvi2019}.
This step is carried out efficiently by applying the Arnoldi method of the ARPACK library \cite{Lehoucq1998}. The optimization is sequentially iterated for all the tensors in the tree network. The whole procedure (sweep) is repeated as long as the total energy does not converge to a minimal value.
As for the single-node optimization of the Tensor Network, we set the Arnoldi algorithm to discard singular values smaller than $10^{-4}$. Then, the convergence of the whole TN algorithm relies on absolute and relative convergence thresholds $\Delta \varepsilon_{abs}=10^{-5}$ and $\Delta \varepsilon_{re\ell}=10^{-5}$ defined respectively as
\begin{align}
  \Delta \varepsilon_{abs}&\equiv\abs{\varepsilon_{n-1}-\varepsilon_{n}}&
  \Delta \varepsilon_{re\ell}&\equiv\abs{\frac{\varepsilon_{n-1}-\varepsilon_{n}}{\varepsilon_{n}}}
\end{align}
for energy values $\varepsilon_{n-1}$ and $\varepsilon_{n}$ arisen from consecutive optimization sweeps $n-1$ and $n$. 
Ultimately, the maximal bond dimension $\chi$ adopted in the reported TN simulations is always obtained by looking at the single-site energy relative convergence of $10^{-4}$ (see \cref{fig_conv_tests}).
\bibliography{SU2.bib}
\end{document}